\documentstyle[multicol,prb,aps,epsf]{revtex}
\renewcommand{\narrowtext}{\begin{multicols}{2}
\global\columnwidth20.5pc\noindent}
\renewcommand{\widetext}{\end{multicols}
\global\columnwidth42.5pc}
\begin{document}
%\narrowtext
\draft

%\twocolumn[\hsize\textwidth\columnwidth\hsize\csname@t%wocolumnfalse%
%\endcsname

\title{Site-selective nuclear magnetic relaxation time 
in a superconducting vortex state}

\author{Mitsuaki Takigawa, Masanori Ichioka and Kazushige Machida}
\address{Department of Physics, Okayama University,
         Okayama 700-8530, Japan}
 \date{\today}

\maketitle

\begin{abstract}
The temperature and field dependences of the site-selective nuclear spin 
relaxation time $T_1$ around vortices are
studied comparatively
both for $s$-wave and 
$d$-wave superconductors, based on the microscopic
Bogoliubov-de Gennes theory. 
Reflecting low energy electronic excitations associated with the 
vortex core, the site selective temperature dependences deviate 
from those of the zero-field case, and $T_1$ becomes faster with 
approaching the vortex core. 
In the core region, $T_1^{-1}$ has a new peak below the superconducting 
transition temperature $T_c$. 
The field dependence of the overall $T_1(T)$ behaviors for
$s$-wave and $d$-wave superconductors is investigated 
and analyzed in terms of the local density of states.
The NMR study by the resonance field dependence may be a new method 
to probe the spatial resolved vortex core structure in various 
conventional and unconventional superconductors. 
\end{abstract}


\narrowtext

%%%%%%%%%%%%%%%%%%%%%%%%%%%%%%%%%%%%%%%%%

\section{Introduction}

Much attention has been focused on vortex matter physics both of 
high $T_c$ cuprates and of conventional superconductors.\cite{machida,sonier}
There are several experimental methods to extract the vortex properties, 
ranging from thermodynamic measurements such as specific heat to various 
transport experiments such as thermalconductivity under field. 
We can divide them into two categories: 
One of the methods such as scanning tunneling microscopy (STM)  probes 
the electronic structure of the vortex, namely, 
low-lying excitations around a core. 
The other method probes the magnetic field distribution in the vortex state,
examplified by small angle neutron diffraction, or
muon spin resonance ($\mu$SR) experiments.
These two methods are complimentary to each other. We definitely need to 
increase the list of the method and further refine the theoretical 
analysis of experimental data to extract useful information 
about the vortex properties.

Among various experimental methods mentioned above the nuclear  
magnetic resonance (NMR) experiment~\cite{Haase} is unique  
because it provides us two kinds of information. Namely,  
the nuclear resonance spectrum reflects the magnetic field  
distribution and the longitudinal relaxation time $T_1$ probes  
electronic excitations in the vortex state through its $T$-dependence. 
The NMR experiment has been playing a vital role in distinguishing 
between $s$-wave and $d$-wave pairing symmetries in this respect, 
i.e. via $T$-dependence of $T_1$. 
The power law $T_1^{-1}\propto T^{3}$ ($T^{5}$) behavior is taken 
as definitive evidence for a line (point) node in the gap 
structure of unconventional superconductors. 
This conclusion comes from a simple power counting of the density of states 
(DOS) at the Fermi level as a 
function of the energy $E$: $N(E)\propto E$ ($E^{3}$) for a 
line (point) node in a bulk superconductor at zero field. 
However, actual NMR experiments are performed under applied fields  
in a mixed state. 
The contribution of the vortex core is inevitably included in their 
data.~\cite{Silbernagel,Ishida} 
Usually, $T_1$ is  measured by selecting the resonance frequency at a 
most intensive signal in the resonance spectrum. 
However, the resonance spectrum reflects information of internal 
magnetic field distribution of the vortex lattice~\cite{Fite,Kung}  
as mentioned above. 
Thus, by choosing the resonance field, we can specify the spatial  
position to detect the NMR signal. 
The signal at the maximum (minimum) cutoff comes from the vortex 
center (the furthest) site. 
The signal at the logarithmic singularity of the resonance field  
comes from the saddle points in the field distribution. 
By studying the position dependence of $T_1$ around vortices through 
the resonance frequency dependence, we can clarify the detail of the  
vortex contribution in the NMR experiments.  
It helps us in the analysis of the standardized procedure extracting  
the gap symmetry. 

Low-lying excitation spectra around a vortex are not fully understood 
both experimentally and theoretically.
The related problems are as follows. 
In the $s$-wave superconductors, the effect of the quantized energy 
level will appear in the quasi-particle 
state.~\cite{CdGM,Caroli,Hayashi,Hayashi2} 
In the $d$-wave case, the low energy state around the vortex core 
extends outside the core due to the node of the superconducting gap, 
and shows the $\sqrt{H}$-like DOS relation ($H$ is an applied 
field).~\cite{Volovik,IchiokaDL,IchiokaDS,Wang,Franz} 
We also need to estimate the quasi-particle transfer between vortices 
(such as the path of the transfer and its amplitude) to study the 
dHvA oscillation or transport phenomena in the mixed 
state.~\cite{IchiokaDL,IchiokaSL} 
The excitation around the core plays a fundamental role in 
determining physical properties of superconductors.
Several recent theories based on the microscopic Bogoliubov-de 
Gennes (BdG) equation investigate  electromagnetic properties of 
mixed state~\cite{wang,janko1} in connection with the low-lying 
vortex excitations.~\cite{CdGM,Wang,Franz,melnikov,janko2,franz2} 
In high $T_c$ cuprates, the existence or non-existence of localized 
core excitations in $d$-wave pairing case is actively debated.
Theoretical study suggested the zero-energy peak in the $d$-wave 
case, instead of the quantized energy level in the $s$-wave 
case.~\cite{IchiokaDS,Wang,Franz} 
On the contrary, the STM 
experiments reported quantized energy level with large gap in 
YBCO (Ref. \onlinecite{Maggio}) and also in BSCCO (Ref. \onlinecite{pan}), 
and  no peak within the 
superconducting gap in BSCCO (Ref. \onlinecite{Renner}). 
A part of reasons of the debate is due to limited experimental 
methods which directly probe the spatially resolved core structure.
So far, the STM was only a method to detect it as mentioned before. 
A large number of thermodynamic or transport measurements probe 
spatially averaged quantities. 

Here we propose a novel spatially resolved means, that is, vortex 
imaging to see electronic excitations associated with a vortex core 
by using NMR, and demonstrate how the $T$-dependence of $T_1$ is 
site-sensitive.
Through this analysis, we are able to produce a spatial image of 
the low-lying excitation spectrum around a core. 
A similar idea of the NMR imaging is actually tested experimentally 
in high $T_c$ materials by Slichter's group~\cite{Slichter,milling} 
and theoretically proposed,\cite{wortis,morr} 
and also in spin-Peierls system CuGeO$_3$ by Horvati\'{c}.~\cite{Horvatic} 

In order to analyze the NMR data and propose 
suitable NMR experiments, we perform a model calculation of 
$T_1$ to demonstrate how careful NMR experiment is valuable. 
The other purpose of this paper is to warn a pitfall when 
obtaining the conclusion of the nodal gap structure from 
the $T$-dependence of $T_1$ by performing  the saddle point NMR.
The position dependence of the NMR signal in the $s$-wave case was 
theoretically studied under some approximations.~\cite{Caroli,Leadon}  
Here, we calculate it microscopically from the wave functions 
obtained by self-consistently solving the BdG equation for the 
extended Hubbard model in the $s$- and $d$-wave cases. 

After giving the formulation of the problem in Sec. \ref{sec2},
we show the results of basic vortex properties
in this formulation in order to yield a coherent physical 
picture and to facilitate the later understanding of our results 
in Sec. \ref{sec3}. 
In next section the $T$-dependence of $T_1$ is analyzed both for
s-wave and d-wave superconductors. 
The field-dependence of $T_1$ is calculated in Sec. \ref{sec5}. 
The final section is devoted to conclusion and discussions.\cite{takigawa}

\section{Formulation}
\label{sec2}

\subsection{Bogoliubov-de Gennes equation on lattice}

The BdG theory and its equation are one
of the most fundamental frameworks in the theory of 
superconductivity.\cite{deGennes} 
In principle, the solution of BdG equation should give all static 
properties of a type II superconductor under an applied field, 
which we concern.
Here we consider the BdG equation defined on a lattice instead of 
continuum space. 
The latter approach is useful when the order parameter is 
described by the s-wave symmetry, but difficult to treat 
non-local higher angular momentum states such as $d$-wave. 
In contrast the lattice BdG theory is relatively easy to treat 
such a case and suitable for treating the $s$-wave 
and $d$-wave cases on an equal footing.

In terms of the eigen-energy $E_\alpha$ and the wave functions 
$u_\alpha({\bf r}_i)$, $v_\alpha({\bf r}_i)$ at $i$-site,
the BdG equation for the extended Hubbard model defined on the 
two-dimensional square lattice is given by
%%%
\begin{equation}
\sum_j
\left( \begin{array}{cc}
K_{i,j} & D_{i,j} \\ D^\dagger_{i,j} & -K^\ast_{i,j}
\end{array} \right)
\left( \begin{array}{c} u_\alpha({\bf r}_j) \\ v_\alpha({\bf r}_j)
\end{array}\right)
=E_\alpha
\left( \begin{array}{c} u_\alpha({\bf r}_i) \\ v_\alpha({\bf r}_i)
\end{array}\right) ,
\label{eq:BdG1}
\end{equation}
%%%
where 
%%%
\begin{equation}
K_{i,j}=-\tilde{t}_{i,j}  - \delta_{i,j} 
\mu,
\label{eq:BdG2}
\end{equation}
%%%
%%%
\begin{equation}
D_{i,j}=\delta_{i,j} U \Delta_{i,i} + \frac{1}{2}V_{i,j} 
\Delta_{i,j} 
\label{eq:BdG3}
\end{equation}
%%%
with 
%%%
\begin{equation}
\tilde{t}_{i,j}=t_{i,j} \exp [ {\rm i}\frac{\pi}{\phi_0}\int_{{\bf 
r}_i}^{{\bf r}_j} {\bf A}({\bf r}) \cdot d{\bf r} ] 
\label{eq:BdG4}
\end{equation}
%%%
and the on-site interaction $U$, the chemical 
potential $\mu$ and the flux quantum $\phi_0$.
The transfer integral $t_{i,j}=t$ and the nearest neighbor 
(NN) interaction $V_{i,j}=V$ 
for the NN site pair ${\bf r}_i$ and ${\bf r}_j$, and otherwise 
$t_{i,j}=V_{i,j}=0$.
The vector potential ${\bf A}({\bf r})=\frac{1}{2}{\bf H}\times{\bf 
r}$ in the symmetric gauge.
The self-consistent condition for the pair potential is
%%%
\begin{equation}
\Delta_{i,j}=-\frac{1}{2}\sum_\alpha u_\alpha({\bf r}_i)
v^\ast_\alpha({\bf r}_j) \tanh(E_\alpha /2T) .
\label{eq:BdGsc}
\end{equation}
%%%
This  BdG equation is self-consistently solved by following 
the numerical method of Wang and MacDonald.\cite{Wang}

In the following we examine comparatively the
$s$-wave and $d$-wave symmetry cases.
The $s$-wave pair potential is given by 
%%%
\begin{equation}
\Delta_s({\bf r}_i)=U\Delta_{i,i}.
\label{eq:sOP1}
\end{equation}
%%%
The $d_{x^2-y^2}$-wave pair potential is given by
%%%
\begin{equation}
\Delta_{d}({\bf r}_i)={V\over4}(\Delta_{\hat{x},i} + \Delta_{-\hat{x},i}
- \Delta_{\hat{y},i} - \Delta_{-\hat{y},i} )
\label{eq:dOP1}
\end{equation}
%%%
with 
%%%
\begin{equation}
\Delta_{\pm\hat{e},i}=\Delta_{i,i \pm \hat{e}}
\exp[{\rm i}\frac{\pi}{\phi_0}
\int_{{\bf r}_i}^{({\bf r}_i+{\bf r}_{i \pm \hat{e}})/2}
{\bf A}({\bf r}) \cdot d{\bf r}].
\label{eq:dOP2}
\end{equation}
%%%

We consider the square vortex lattice case where nearest neighbor 
vortex is located at the $45^\circ$ direction from the $a$ axis.
This vortex lattice configuration is suggested for $d$-wave 
superconductors, or $s$-wave superconductors with fourfold symmetric 
Fermi surface.~\cite{IchiokaDL,square-vortex,Won}
The unit cell in our calculation is the square area of $N_r^2 $ sites 
where two vortices are accommodated.
Then, the magnetic field is determined as
%%%
\begin{equation}
H={2 \phi_0 \over(c N_r)^2} 
\label{eq:field}
\end{equation}
%%%
where $c$ is the atomic lattice constant. 
For typical high T$_c$ cuprates, $H_{N_r=30}$
corresponds to an order of 15 Tesla.
We consider the area of $N_k^2$  unit cells.
By introducing the quasi-momentum of the magnetic Bloch state,
%%%
\begin{equation}
{\bf k}={2 \pi \over c N_r N_k}(l_x,l_y), \qquad 
(l_x,l_y=1,\cdots, N_k)  
\label{eq:kpoint}
\end{equation}
%%%
we set
%%%
\begin{equation}
u_\alpha({\bf r})=\tilde{u}_\alpha({\bf r}) 
{\rm e}^{{\rm i} {\bf k}\cdot{\bf r}}, \qquad
v_\alpha({\bf r})=\tilde{v}_\alpha({\bf r}) 
{\rm e}^{{\rm i} {\bf k}\cdot{\bf r}}.
\label{eq:u-function}
\label{eq:v-function}
\end{equation}
%%%
We solve Eq. (\ref{eq:BdG1}) within a unit cell under the given 
${\bf k}$.
Then, $\alpha$ is labeled by ${\bf k}$ and the eigen-values obtained 
by this calculation within a unit cell.

\subsection{Boundary condition}

We impose the periodic boundary condition 
given by the symmetry for  the 
translation 
%%%
\begin{equation}
{\bf R}=m{\bf u}_1 +n{\bf u}_2 
\label{eq:R-function}
\end{equation}
%%% 
with $m$ and $n$ being 
integers, and
${\bf u}_1$ and ${\bf u}_2$ are unit vectors of the vortex 
lattice, i.e., 
%%%
\begin{equation}
\tilde{u}_\alpha({\bf r}+{\bf 
R})=\tilde{u}_\alpha({\bf r}) {\rm e}^{i\chi({\bf r},{\bf R})/2}, 
\label{eq:tildeu-function}
\end{equation}
%%% 
%%%
\begin{equation}
\tilde{v}_\alpha({\bf r}+{\bf R})=\tilde{v}_\alpha({\bf r}) 
{\rm e}^{-i\chi({\bf r},{\bf R})/2}. 
\label{eq:tildev-function}
\end{equation}
%%% 
Here, 
%%%
\begin{equation}
\chi({\bf r},{\bf R})
= -\frac{2\pi}{\phi_0}{\bf A}({\bf R})\cdot{\bf r}
- \pi m n + \frac{2 \pi}{\phi_0}
({\bf H}\times {\bf r}_0)\cdot{\bf R}
\label{eq:chi-function}
\end{equation}
%%%
in the symmetric gauge when the vortex center is located at
%%%
\begin{equation}
{\bf r}={\bf r}_0+\frac{1}{2}({\bf u}_1+{\bf u}_2).
\label{eq:R}
\end{equation}
%%%
The phase factor~\cite{Ozaki} in Eq. (\ref{eq:dOP2}) is needed 
to satisfy the translational relation 
%%%
\begin{equation}
\Delta_{d}({\bf r}+{\bf R})
=\Delta_{d}({\bf r}){\rm e}^{{\rm i}\chi({\bf r},{\bf R})}.
\label{eq:dOP12}
\end{equation}
%%%

\subsection{Thermal Green function}

In order to calculate various physical quantities,
we must construct the Green functions from $E_\alpha$, $u_\alpha({\bf 
r})$, $v_\alpha({\bf r})$ defined as
%%%
\begin{eqnarray} &&
\hat{g}(x,x') 
\equiv \left(
 \begin{array}{cc} 
g_{11}(x,x') & g_{12}(x,x') \\ g_{21}(x,x') & g_{22}(x,x') 
 \end{array}  \right) 
\nonumber  \\ &&
\equiv \left( 
\begin{array}{cc}
-\left< T_{\tau}[\hat{\psi}_{\uparrow}(x)
       \hat{\psi}_{\uparrow}^{\dagger}(x')] \right> & 
-\left< T_{\tau}[\hat{\psi}_{\uparrow}(x)
       \hat{\psi}_{\downarrow}(x')] \right> \\ 
-\left<T_{\tau}[\hat{\psi}_{\downarrow}^{\dagger}(x)
       \hat{\psi}_{\uparrow}^{\dagger}(x')] \right> & 
-\left<T_{\tau}[\hat{\psi}_{\downarrow}^{\dagger}(x)
       \hat{\psi}_{\downarrow}(x')]\right> \end{array} 
\right).
\label{eq:green}
\end{eqnarray}
%%%
with $x\equiv({\bf x},\tau)$.  
After the Fourier transformation of $\tau$ as
%%%
\begin{eqnarray}
\hat{g}(x,x') 
= T\sum_{\omega_n} e^{-i\omega_n(\tau-\tau')}
\hat{g}({\bf{x}},{\bf{x'}},\omega_n), 
\end{eqnarray} 
 %%%
the thermal Green functions with the 
Fermionic imaginary frequency $\omega_n=2\pi T(n+{1\over2})$
are written as
%%%
\begin{eqnarray} 
g_{11}({\bf{x}},{\bf{x'}},\omega_n) &=& \sum_{\alpha}
\frac{u_{\alpha}({\bf{x}})u_{\alpha}^{*}({\bf{x'}})}
{i\omega_n-E_{\alpha}} 
\label{eq:green2-1}\\ 
g_{12}({\bf{x}},{\bf{x'}},\omega_n) &=& \sum_{\alpha}
\frac{u_{\alpha}({\bf{x}})v_{\alpha}^{*}({\bf{x'}})} 
{i\omega_n-E_{\alpha}} \\ 
g_{21}({\bf{x}},{\bf{x'}},\omega_n) &=& \sum_{\alpha}
\frac{v_{\alpha}({\bf{x}})u_{\alpha}^{*}({\bf{x'}})} 
{i\omega_n-E_{\alpha}} \\
g_{22}({\bf{x}},{\bf{x'}},\omega_n) &=& \sum_{\alpha}
\frac{v_{\alpha}({\bf{x}})v_{\alpha}^{*}({\bf{x'}})} 
{i\omega_n-E_{\alpha}}. 
\label{eq:green2-4}
\end{eqnarray}
%%%
The derivation of this form of the thermal Green functions 
is given in Appendix \ref{secA}. 

\subsection{Local density of states}

To understand the behavior of the position-dependent
$T_1({\bf r})$, we also consider the 
local density of states (LDOS) at $\bf{r}$. 
This is evaluated by using the thermal Green functions as 
%%%
\begin{equation}
N_{\uparrow}(E,{\bf{r}})=
-{1\over \pi}{\rm Im}g_{11}({\bf{r}},{\bf{r}},i\omega\rightarrow E+i\eta)
\end{equation} 
%%%
for the up-spin electron contributions, and 
%%%
\begin{equation}
N_{\downarrow}(E,{\bf{r}})
={1\over \pi}{\rm Im}g_{22}({\bf{r}},{\bf{r}},i\omega\rightarrow E+i\eta)
\label{eq:ldosdown}
\end{equation}
%%%
for the down-spin electron contributions. 
Then, the LDOS is given by 
%%%
\begin{eqnarray} &&
N(E,{\bf{r}}) 
= N_{\uparrow}(E,{\bf{r}}) + N_{\downarrow}(E,{\bf{r}}) 
\nonumber \\ && 
=\sum_\alpha \{ |u_\alpha({\bf{r}})|^2\delta(E-E_\alpha)
+|v_\alpha({\bf{r}})|^2\delta(E+E_\alpha) \} . 
\label{eq:LDOS}
\end{eqnarray} 
%%%
For finite temperatures, the $\delta$-functions in eq. (\ref{eq:LDOS}) 
are replaced by the derivative $f'(E)$ of the Fermi distribution 
function $f(E)$:
%%%
\begin{eqnarray} &&
N(E,{\bf r})
\nonumber \\ &&
=-\sum_\alpha [|u_\alpha ({\bf r})|^2f'(E_\alpha -E)
+ |v_\alpha ({\bf r})|^2f'(E_\alpha +E)].
\label{eq:STM}
\end{eqnarray}
%%%
This finite temperature LDOS  corresponds to the differential 
tunnel conductance of STM experiments.

\subsection{Nuclear relaxation time}

We now evaluate the spin-spin correlation 
function $\chi_{-,+}({\bf r},{\bf r}',i \Omega_n)$.\cite{Leadon} 
Its derivation is given in Appendix \ref{secB}.
We obtain the nuclear spin relaxation rate by using Eq. (\ref{eq:chi}),
%%%
\begin{eqnarray}
R({\bf r},{\bf r}') &=&
{\rm Im}\chi_{-,+}({\bf r},{\bf r}',
i \Omega_n \rightarrow \Omega + {\rm i}\eta)/(\Omega/T)|_{\Omega 
\rightarrow 0}
\nonumber \\
&=&
 -\sum_{\alpha,\beta} u_\alpha({\bf r})u^\ast_{\beta}({\bf r})
[ u_\alpha({\bf r}')u^\ast_{\beta}({\bf r}')
 +v_\alpha({\bf r}')v^\ast_{\beta}({\bf r}') ]
\nonumber \\ &&
\times \pi T f'(E_\alpha) \delta(E_\alpha - E_{\beta}).
\label{eq:R1}
\end{eqnarray}
%%%
We consider the case ${\bf r}={\bf r}'$ by assuming that 
the nuclear relaxation event occurs locally such as in Cu-site 
of high $T_c$ cuprates.
Then, the ${\bf r}$-dependent relaxation time is given by 
%%%
\begin{eqnarray}
T_1({\bf 
r})=1/R({\bf r},{\bf r}).
\label{eq:T1}
\end{eqnarray}
%%%
In Eq. (\ref{eq:R1}), we use $\delta(x)=\pi^{-1} {\rm Im}(x-{\rm 
i}\eta)^{-1}$ to consider the discrete energy level of the finite 
size calculation. 
We typically use $\eta=0.01t$.

% ###### FIG. 1 start ###################################################
\vspace{5.5mm}
\begin{figure}[tbp]
\begin{center}
\leavevmode
\epsfxsize=60mm
\epsfbox{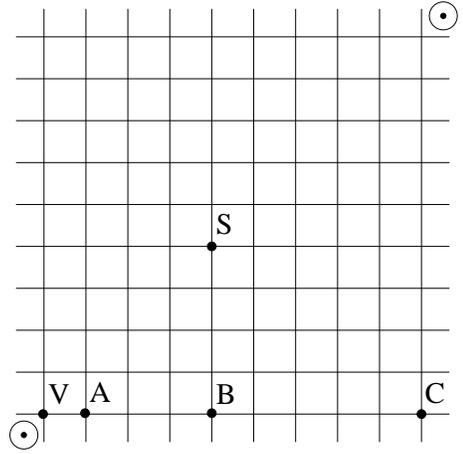}
\end{center}
\caption{
A quarter region of the vortex unit cell which contains two vortices.
Position of the sites V, A, B, C, and S in the square vortex 
lattice is indicated, 
where the nearest neighbor vortex is located in the $45^\circ$ 
direction from the $a$-axis.
The vortex center is shown by $\odot$. 
The solid lines show the square atomic lattice.
}
\label{fig:unit}
\end{figure}
% ###### FIG. 1  end ####################################################
\widetext
\narrowtext

\subsection{Parameters used and site assignment}

The following parameter values are chosen:
The average electron density per site $\sim 0.9$ by 
adjusting appropriately the chemical potential value $\mu$.
We normalize all the energy scales by the
transfer integral $t$.
For the $s$-wave case $U=-2.32t$ and $V=0$.
The resulting order parameter $\Delta_0/t=0.5$ at $T$=0 and $H$=0. 
For the $d$-wave case $U=0$ and $V=-4.2t$. Then $\Delta_0/t=1.0$.
The vortex unit cell is shown in Fig.\ref{fig:unit}, 
where the selective sites (V,A,B,C, and S-sites) to consider 
$N(E,{\bf r})$ and $T_1({\bf r})$ are also indicated.
The magnetic field $H$ is denoted by the unit cell size $N_r$ as $H_{N_r}$.
Thus as $N_r$ increases, $H$ itself decreases.

\section{Basic vortex properties}
\label{sec3} 

In order to facilitate the understanding of the 
$T_1({\bf r})$-behaviors in the later section, 
we show some of the basic properties of vortex lattice. 
The following results coincide basically with those in the previous 
our own  calculations either in quasi-classical theory for 
the vortex lattice of $s$- and $d$-wave 
cases,~\cite{IchiokaDL,IchiokaDS,Ichioka1,Ichioka2,Ichioka4,Ichioka5} 
and in BdG for isolated vortex of the $s$-wave case~\cite{Hayashi,Hayashi2}
and also with the preceding work by Wang and MacDonald.~\cite{wang} 

\subsection{Order parameter profiles}

The pair potential or the order parameter $\Delta({\bf r})$ vanishes 
at the vortex center. $|\Delta({\bf r})|$ exhibits the Friedel 
oscillations around the core both for the $s$-wave and $d$-wave cases
whose period $\sim 1/\/k_F$ ($k_F$ is the Fermi wave number). 
The amplitude of this quantum oscillation increases as the attractive 
interactions $|U|$ and $|V|$ become large because the quantum effects 
are  enhanced when $\Delta_0/E_F$ increases 
($E_{\rm F}$ is the Fermi energy). 
These characteristics coincide with those in the $s$-wave case 
in our previous study.~\cite{Hayashi} 

As for the temperature dependence of the order parameter, 
the vortex core radius shrinks with decreasing 
$T$ by the Kramer-Pesch effect.~\cite{IchiokaDS,Kramer} 
We confirm it for the $d$-wave case too.
The shrinkage is saturated at a low temperature both in the $s$- 
and $d$-wave cases.
There, the structure of $\Delta_s({\bf r})$ and $\Delta_d({\bf r})$ 
is almost independent of $T$.
This is a quantum-limit effect which occurs for $T/T_c < 
\Delta_0/E_{\rm F}$.~\cite{Hayashi} 
We calculate the low temperature behavior of $T_1({\bf r})$ by using 
the saturated pair potential.
At higher temperatures, we calculate $T_1({\bf r})$ by using the 
self-consistently obtained pair potential at each $T$.

% ###### FIG. 2 start ###################################################
\widetext
\vspace{5.5mm}
\begin{figure}[tbp]
\mbox{
\epsfxsize=80mm
\epsfbox{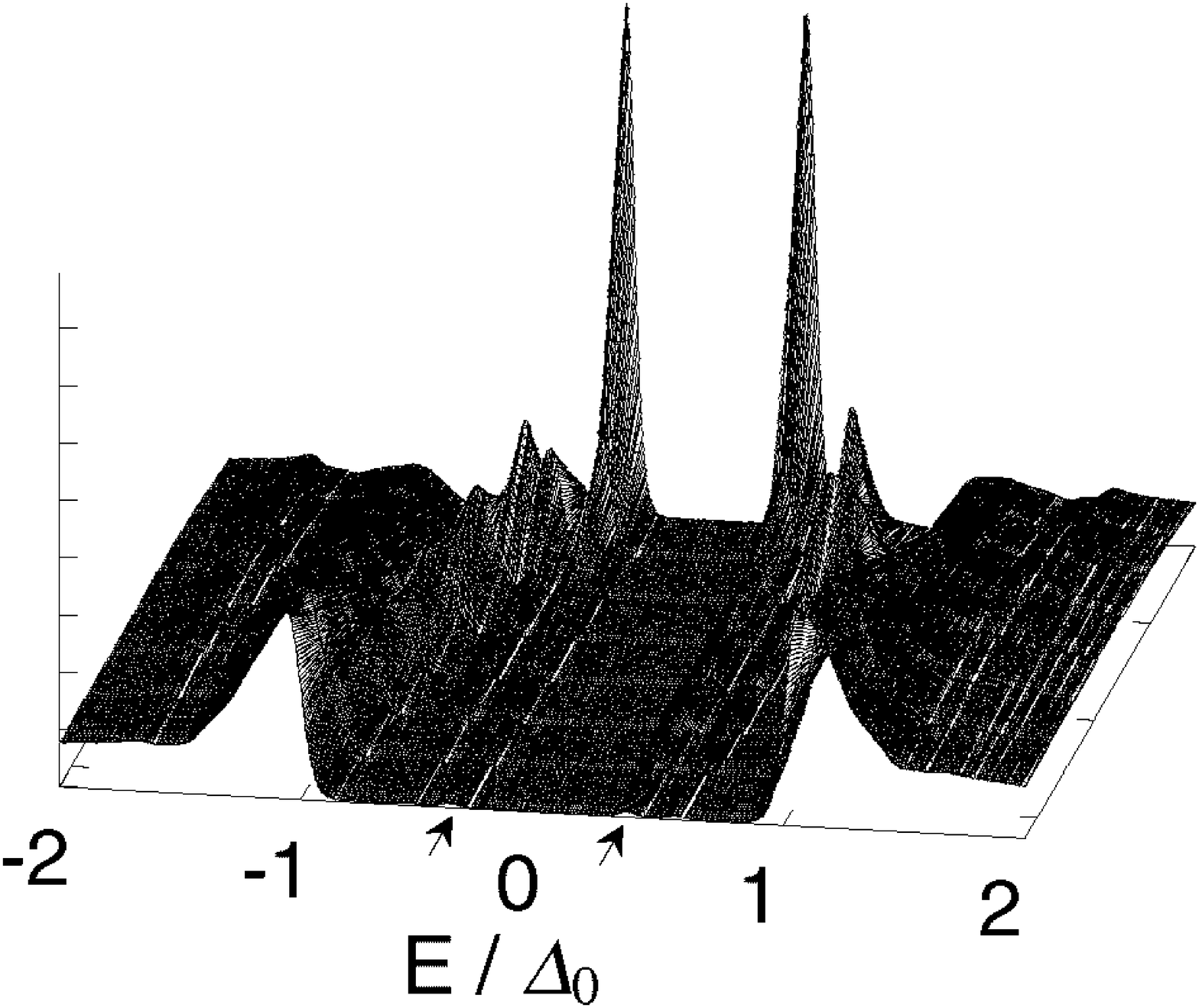}
\epsfxsize=80mm
\epsfbox{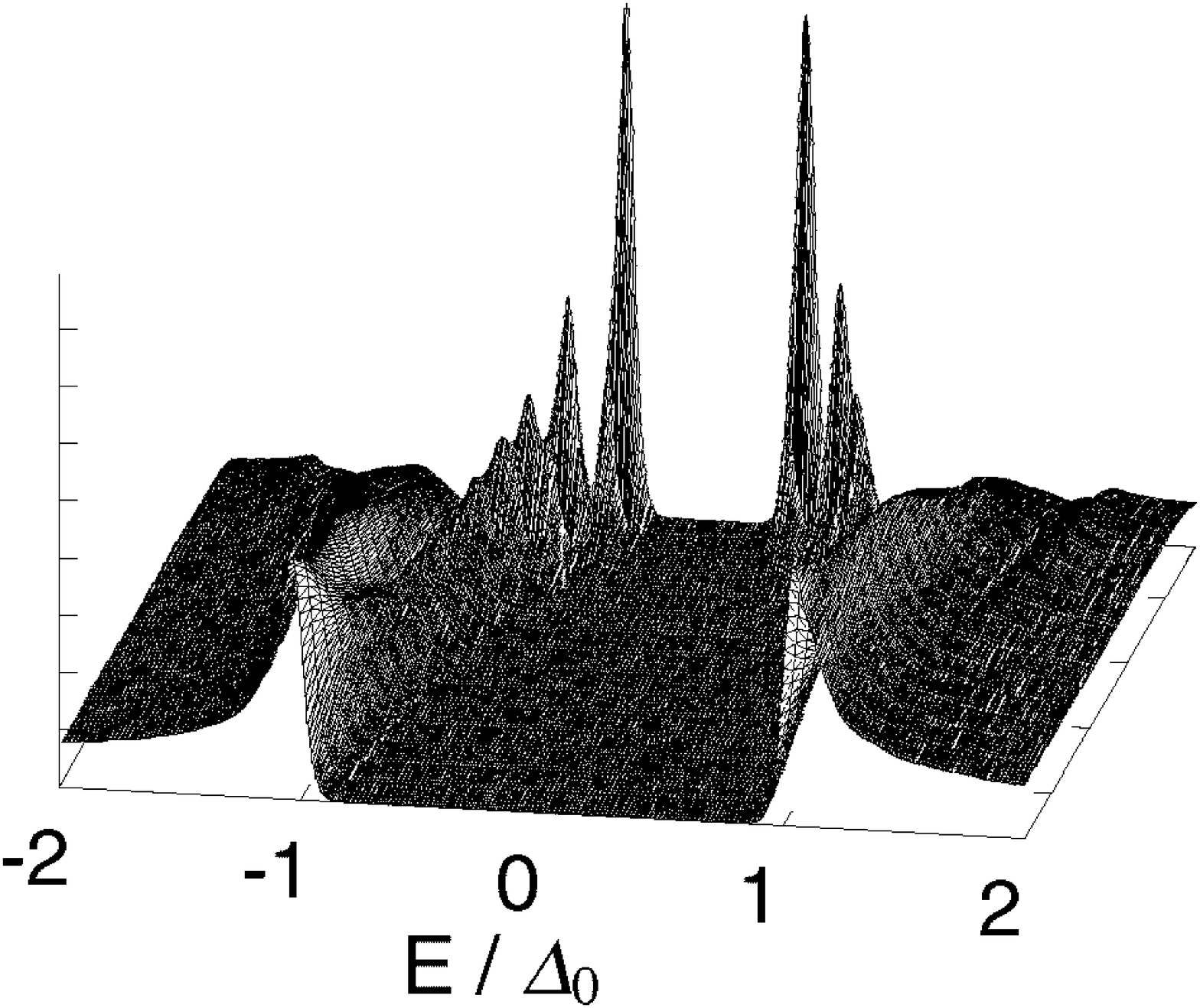}
}
\caption{
(a) 
Local density of states $N({\bf r},E)$ for the $s$-wave case 
at $T=0$ and $H_{32}$ along the NN direction
(V $\rightarrow$ S). The backside (front) corresponds to V (S)-site 
defined in Fig. \ref{fig:unit}. 
(b) 
Local density of states $N({\bf r},E)$ along the NNN direction
(V $\rightarrow$ C). Note that the foot 
at the lowest levels at $E/\Delta_0\sim \pm 0.28$ extends outward 
(arrows) along the NN direction.
}
\label{fig:sdos}
\end{figure}
\narrowtext
% ###### FIG. 2  end ####################################################

\subsection{Local density of states around a core}

\begin{figure}[tbp]
\epsfxsize=80mm
\epsfbox{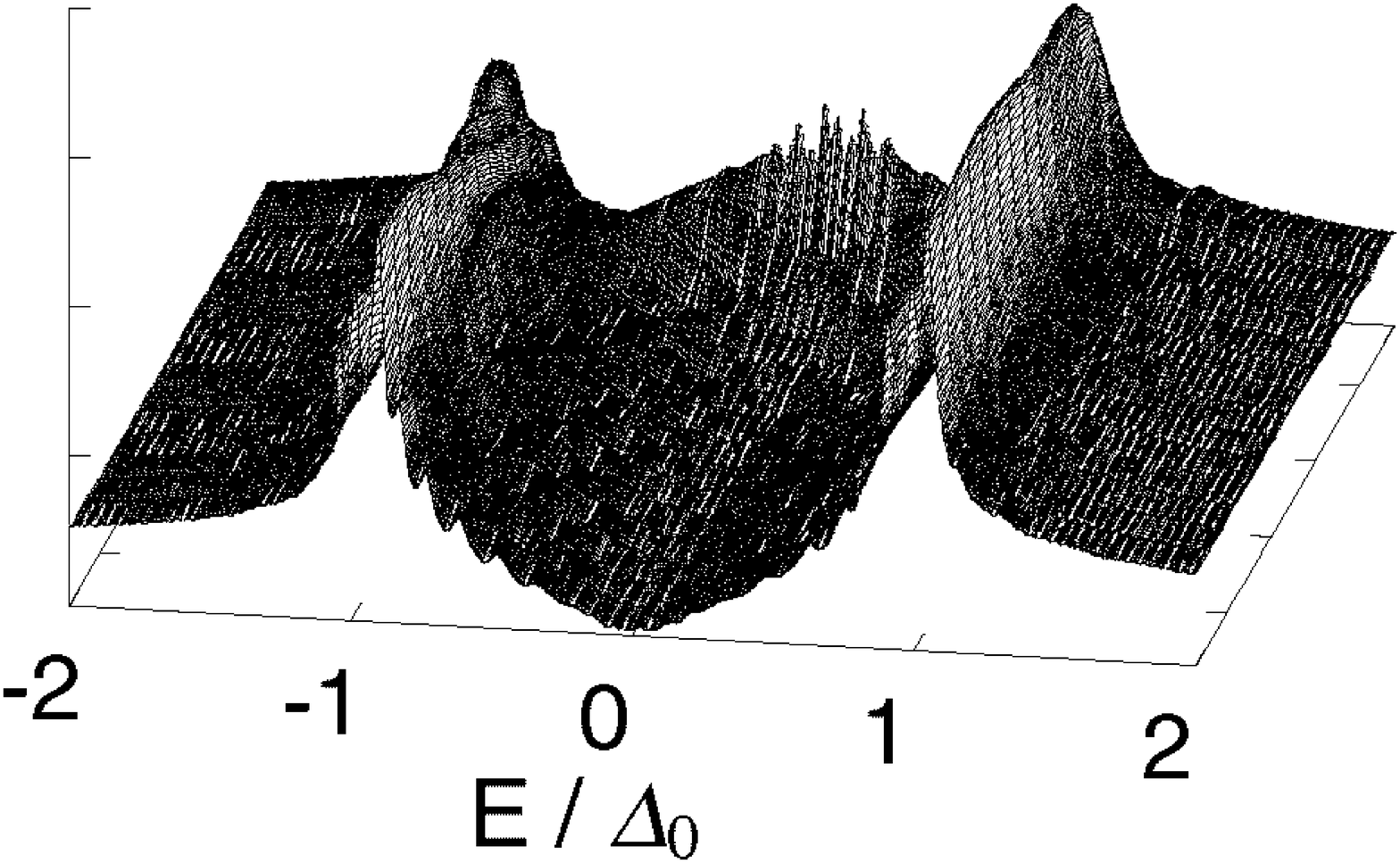}
\vspace{5mm}
\caption{
Local density of states $N({\bf r},E)$ at $T=0$ and $H_{32}$ 
along the NNN direction (V $\rightarrow$ C) for the $d$-wave case.
The backside (front) corresponds to V (C)-site defined 
in Fig. \protect\ref{fig:unit}. 
The spectral evolution along the NN direction, 
which is not shown, shows similar behavior to this figure. 
}
\label{fig:ddos}
\end{figure}

The LDOS is displayed as a function of the 
spatial position for the $s$-wave case in Fig. \ref{fig:sdos} 
and for the $d$-wave case in Fig. \ref{fig:ddos} at $H_{32}$. 
The spectral evolutions in Fig. \ref{fig:sdos} are shown 
for two different paths; along the nearest neighbor (NN)
direction [Fig. \ref{fig:sdos}(a)] and the next nearest neighbor 
(NNN) direction [Fig. \ref{fig:sdos}(b)].
The low-lying excitations at the core, which correspond to 
the prominent peak structure, are characterized by the angular 
momentum (see for details Ref.\onlinecite{Hayashi}).
The higher and higher angular momentum states situated 
at the higher energy are activated successively when going outward, 
forming the characteristic spectral evolution in Fig. \ref{fig:sdos}. 
The lowest excited state which is seen as the highest peak extends 
towards the NN direction  rather than the NNN direction
as indicated by arrows in Fig. \ref{fig:sdos}(a).

\begin{figure}[tbp]
\epsfxsize=80mm
\epsfbox{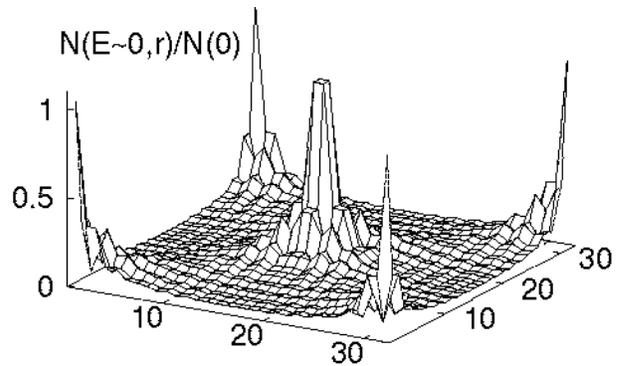}
\caption{
Topographic view of the local density of states
at the Fermi level normalized by the normal state value: 
$N(E=0,{\bf r})/N(0)$ for the $d$-wave case at $H_{32}$.
One unit cell (32$\times$32 atomic sites) is shown.
The vortices are located at the center and corners of the figure.
}
\label{fig:ddos3232}
\end{figure}

The LDOS for the $d$-wave case is shown 
in Fig. \ref{fig:ddos}. 
Since there is no discretized bound state around the core in this case, 
only the broad resonance peak centered at $E=0$ is seen 
at the core, which evolves smoothly. 
At the farthest site 
the spectrum exhibits the known bulk behavior expected 
in the $d$-wave superconductor, characterized by $E$-linear 
behavior due to the line node.
In contrast with the $s$-wave case, there is no clear directional 
dependence in the spectral evolution in this scale (also see 
Fig. \ref{fig:ddos3232}).
\widetext
\narrowtext
% ###### FIG. 3 start ###################################################
\vspace{1mm}
% ###### FIG. 3  end ####################################################

% ###### FIG. 4 start ###################################################
\begin{figure}[tbp]
\vspace{5.5mm}
\begin{center}
\leavevmode
\epsfxsize=60mm
\epsfbox{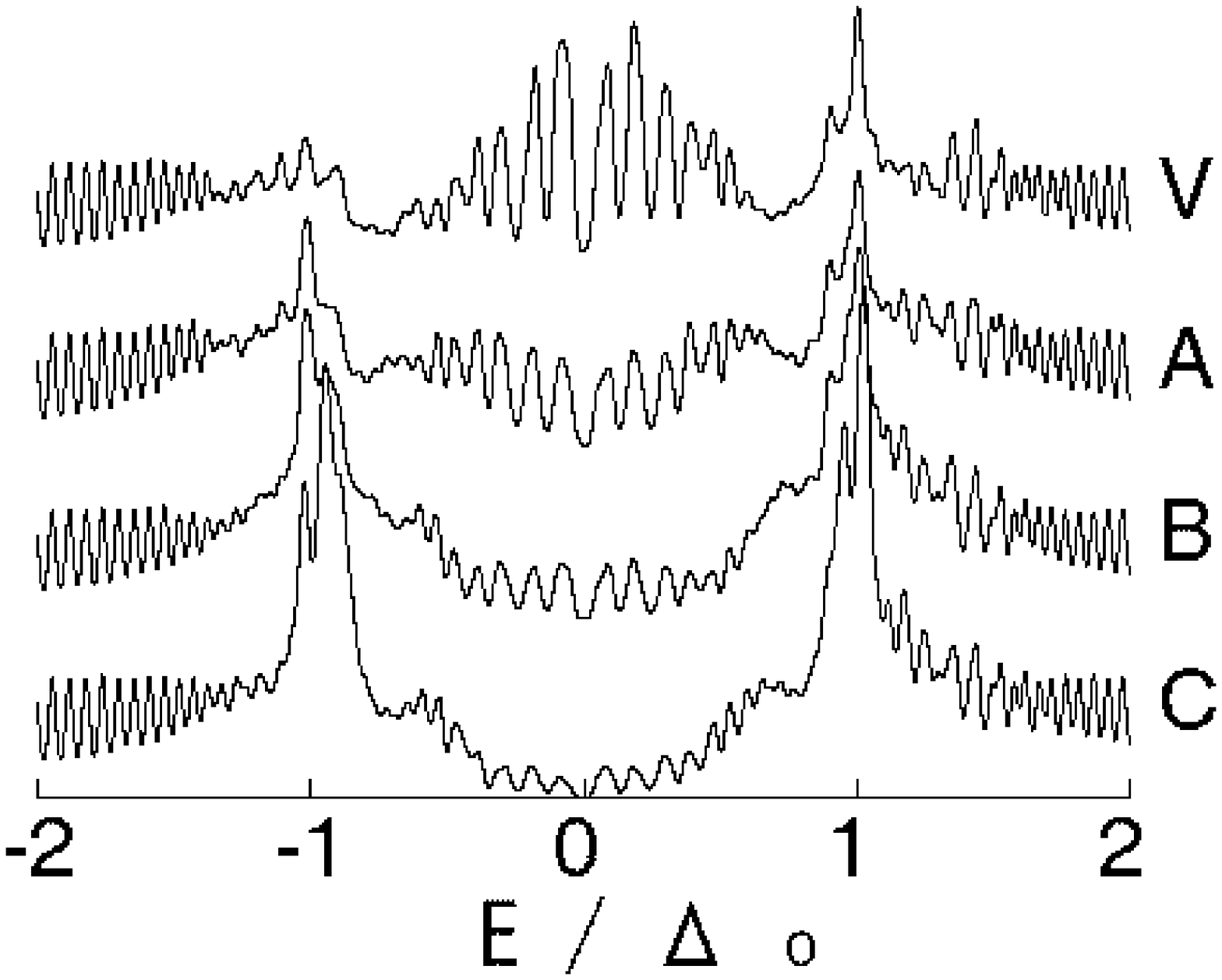}
\end{center}
\caption{
Local density of states $N({\bf r},E)$ at $T=0$ and $H_{32}$ 
along the NNN direction (V $\rightarrow$ C) for the $d$-wave case.
The backside (front) corresponds to V (C)-site defined 
in Fig. \protect\ref{fig:unit}. 
The spectral evolution along the NN direction, 
which is not shown, shows similar behavior to this figure. 
}
\label{fig:ddoshigh}
\end{figure}
% ###### FIG. 4  end ####################################################
\clearpage
% ###### FIG. 5 start ###################################################
\widetext
\vspace{5.5mm}
\begin{figure}[tbp]
\begin{center}
\leavevmode
\mbox{
\epsfxsize=60mm
\epsfbox{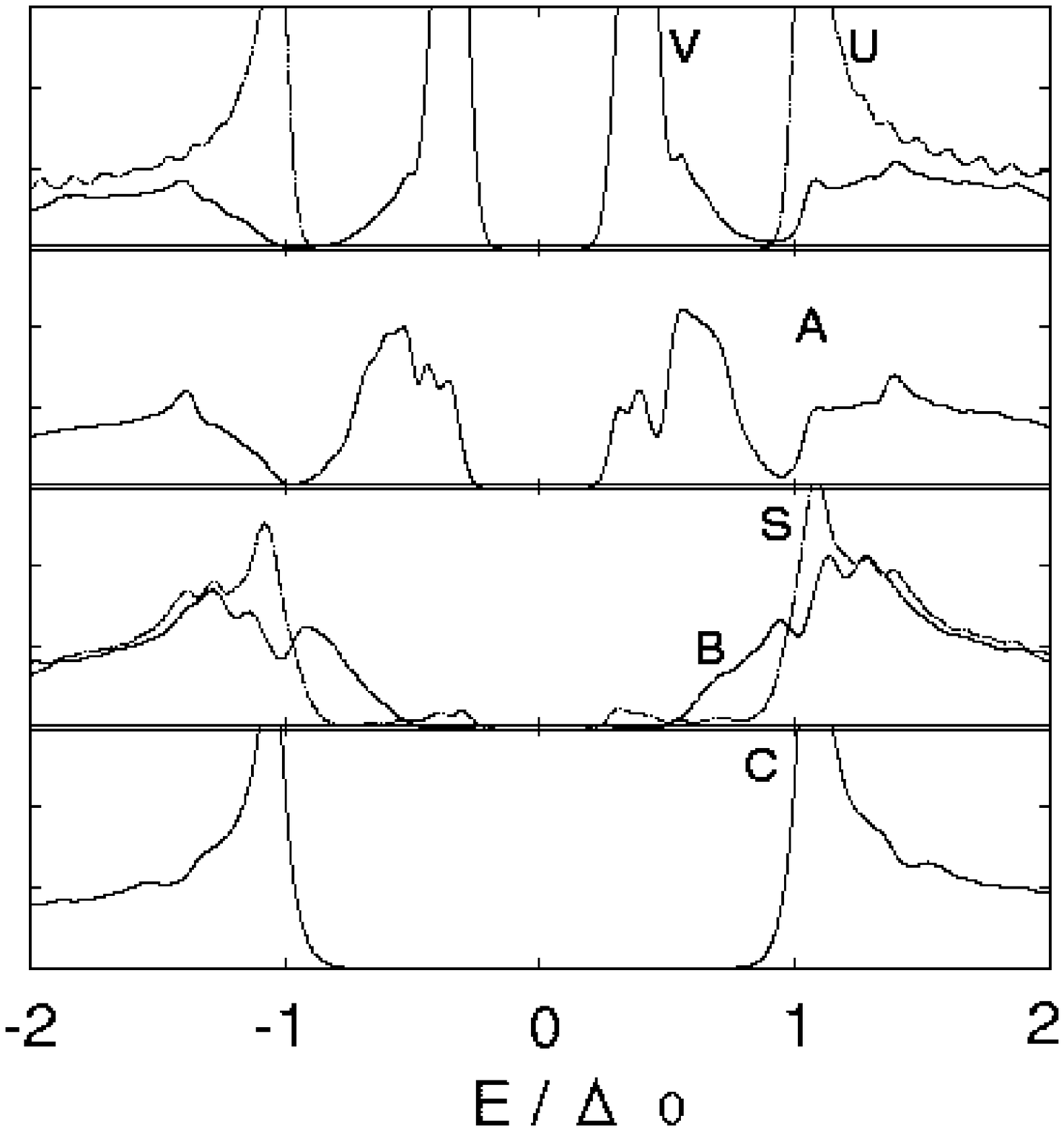}
\epsfxsize=60.5mm
\epsfbox{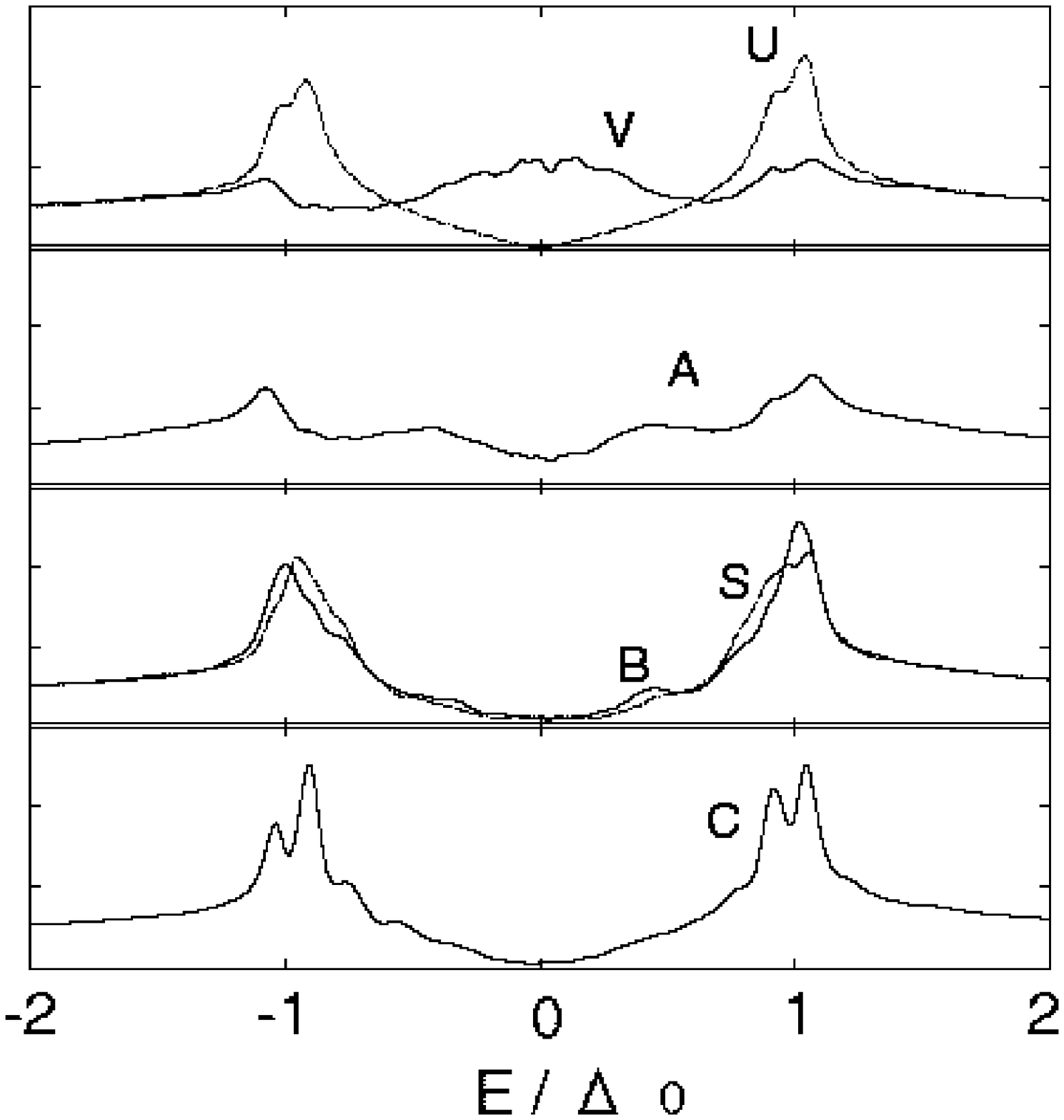}
}
\end{center}
\caption{
Local density of states $N({\bf r},E)$ at $T=0$ and $H_{32}$ 
along the NNN direction (V $\rightarrow$ C) for the $d$-wave case.
The backside (front) corresponds to V (C)-site defined 
in Fig. \protect\ref{fig:unit}. 
The spectral evolution along the NN direction, 
which is not shown, shows similar behavior to this figure. 
}
\label{fig:LDOS}
\end{figure}
\narrowtext
% ###### FIG. 5  end ####################################################

The spectral evolution in a high field $H_{12}$ is depicted for 
the $d$-wave case in Fig. \ref{fig:ddoshigh}. 
The quantum oscillations due to the Landau band 
quantization becomes evident at a high field. 
We see  rapid oscillations outside the main gap ($|E/\Delta_0|>1.0$) 
and slow oscillations inside the main gap ($|E/\Delta_0|<1.0$). 
Note that the split features around $E\sim 0$
at the V-site in Fig. \ref{fig:ddoshigh} is not due
to the localized bound state formation.
In order to further facilitate the understanding of 
the $T$-dependence of the site-dependent $T_1({\bf r})$ behavior, 
we look into the spectral evolutions 
for some more details at $H_{20}$ where we calculate $T_1({\bf r})$ 
in the next section.
The LDOS around the vortex for the $s$-wave case is shown 
in Fig. \ref{fig:LDOS}(a), and the $d$-wave case is shown in 
Fig. \ref{fig:LDOS}(b).
In $N(E,{\bf r})$ at the vortex center (the V-site), the gap edge at 
$\Delta_0$ in the zero-field case (dotted line U)  is smeared, and 
low-energy peaks of the vortex core state appear. 
In the $s$-wave case, we see some peaks above the small gap 
$\Delta_1$ ($\sim \Delta_0^2/E_{\rm F})$.
It is due to the quantization of the energy level in the $s$-wave 
case. 
In the $d$-wave case, the core state shows zero-energy peak instead 
of the split peaks in the $s$-wave case.~\cite{Wang} 
There is no small gap. 
The weight of the low-energy states is decreased with going away from 
the vortex center (V$\rightarrow$A$\rightarrow$B$\rightarrow$C).
Far from the vortex, $N(E,{\bf r})$ is reduced to the DOS of the 
zero-field case.
But, small weight of the low-energy state extending from the vortex 
core remains there.
It is noted that the weight of the low-energy state at the S-site is 
larger than that of the B-site in the $s$-wave case, while the S-site 
is farther from the vortex center [see lines for the S- and B-sites 
in Fig \ref{fig:LDOS}(a)]. 
It is due to the vortex lattice effect.
The quasi-particle transfer between vortices occurs along the line 
connecting NN vortices (i.e., near the S-site).
This is also seen from Fig. \ref{fig:sdos}(a) indicated by arrows.
\\ \ To obtain another perspectives of 
the LDOS image in the $d$-wave case 
we plot the spatial distribution of $N(E=0,{\bf r})$ at $H_{32}$ 
in Fig. \ref{fig:ddos3232}, which directly relates to the $T_1({\bf r})$
behavior as discussed later.
Notice that everywhere $N(E=0,{\bf r})\neq0$.
The LDOS $N(E=0,{\bf r})$ is largest at the core site and extends 
towards the NN direction. 
This shows the characteristic quasi-particle trajectory 
in the $d$ wave case. 
The NMR experiment can image this low-energy excitation as is 
explained later.
The detailed expositions of the quasi-particle trajectories 
in the $s$-wave~\cite{IchiokaDS,IchiokaSL} and $d$-wave~\cite{Ichioka2} 
are given previously.
% ###### FIG. 6 start ###################################################
\vspace{5.5mm}
% ###### FIG. 6 end ####################################################

% ###### FIG. 7 start ###################################################
\widetext
\vspace{5.5mm}
\begin{figure}[tbp]
\begin{center}
\leavevmode
\mbox{
\epsfxsize=60mm
\epsfbox{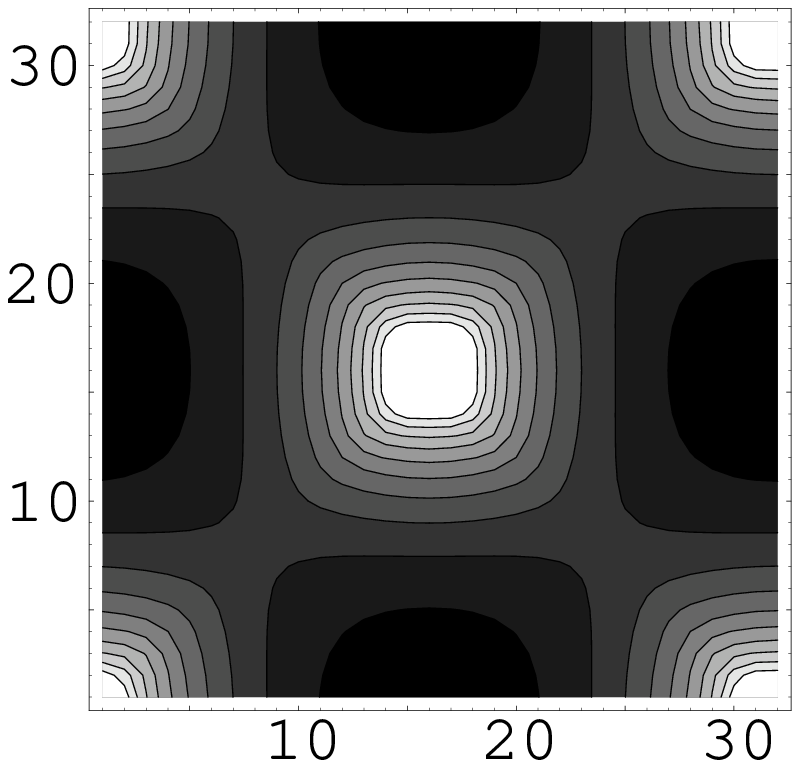}
\epsfxsize=60mm
\epsfbox{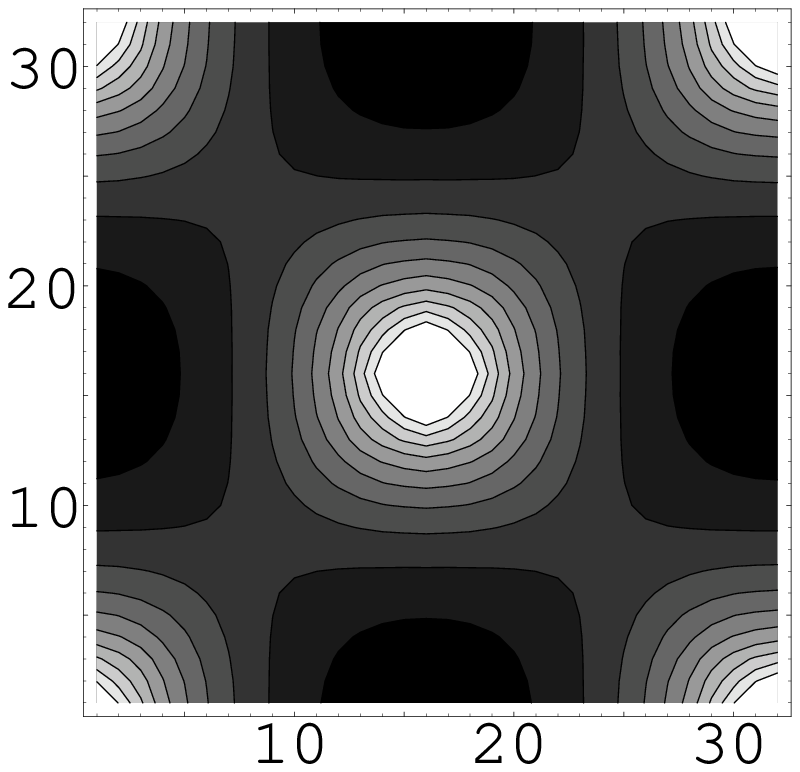}
}
\end{center}
%\vspace{6.0mm}
\caption{
Topographic view of the local density of states
at the Fermi level normalized by the normal state value: 
$N(E=0,{\bf r})/N(0)$ for the $d$-wave case at $H_{32}$.
One unit cell (32$\times$32 atomic sites) is shown.
The vortices are located at the center and corners of the figure.
}
\label{fig:H3232}
\end{figure}
\narrowtext
% ###### FIG. 7 end ####################################################

% ###### FIG. 8 start ###################################################
\widetext
\vspace{5.5mm}
\begin{figure}[tbp]
\begin{center}
\leavevmode
\mbox{
\epsfxsize=60mm
\epsfbox{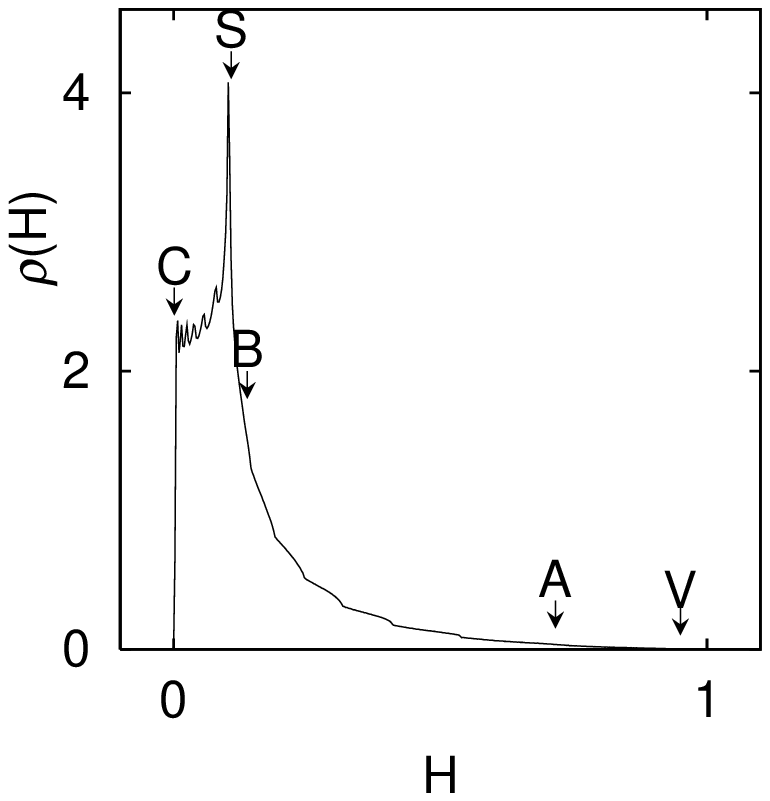}
\epsfxsize=60mm
\epsfbox{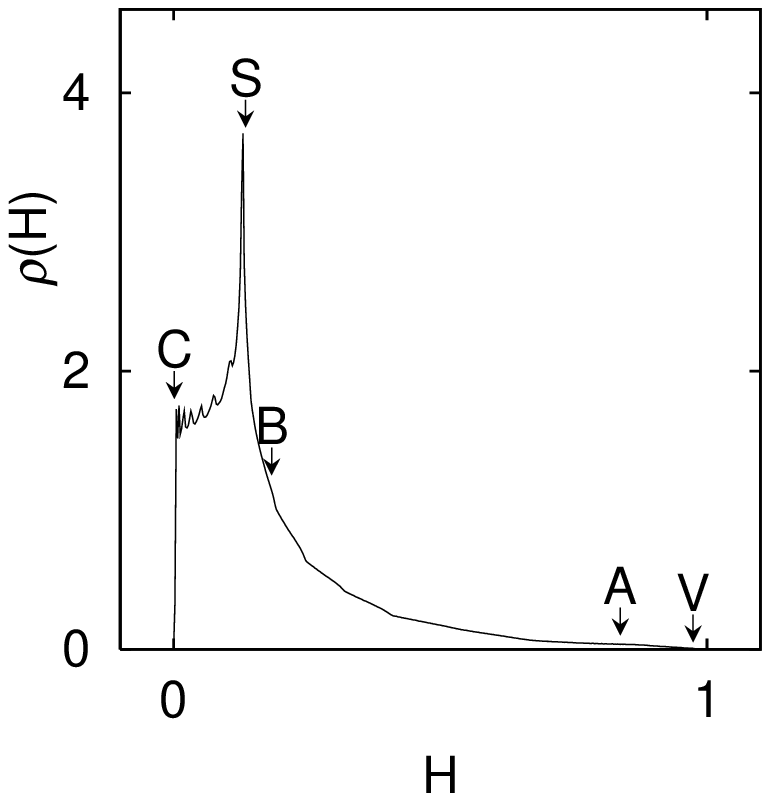}
}
\end{center}
%\vspace{6.0mm}
\caption{
Histograms $\rho(H)$ of the amplitudes of the magnetic field for
the $s$-wave case (a) and the $d$-wave case (b).
This should correspond to the NMR line shape.
The letters in the figures denote the sites defined 
in Fig. \ref{fig:unit}. 
}
\label{fig:lineshape}
\end{figure}
\narrowtext
% ###### FIG. 8 end ####################################################

\subsection{Field distribution in vortex state}

In order to determine the one-to-one correspondence between 
the site and the magnetic field, we calculate the field 
distributions for the $s$-wave and $d$-wave cases. 
The magnetic field is evaluated via the current distribution 
through the Maxwell equation: 
%%%
\begin{equation}
\nabla\times {\bf H}({\bf r})={4\pi\over c}{\bf j}({\bf r}), 
\label{eq:maxwell}
\end{equation}
%%%
where the current ${\bf j}({\bf r})$ is calculated as  
%%%
\begin{eqnarray} 
j_{\hat{e}}({\bf r}_i)
&=& 
2 |e| c {\rm Im}\{ {\tilde t}_{i+\hat{e},i} 
\sum_\sigma \langle \hat{\Psi}^\dagger_{\sigma}({\bf r}_{i+{\hat e}}) 
\hat{\Psi}_{\sigma}({\bf r}_{i}) \rangle \} 
\\ 
&=& 
2 |e| c {\rm Im}\{ {\tilde t}_{i+{\hat e},i} 
\sum_\alpha [ u^\ast_\alpha({\bf r}_{i+{\hat e}})u_\alpha({\bf r}_{i}) 
f(E_\alpha) \nonumber \\ &&
+ v_\alpha({\bf r}_{i+{\hat e}})v^\ast_\alpha({\bf r}_{i}) 
(1-f(E_\alpha)) ] \}
\label{eq:current}
\end{eqnarray}
%%%
for the $\hat{e}$-direction bond ($\hat{e}=\pm\hat{x}$, $\pm\hat{y}$) 
at the site ${\bf r}_i$. 

Figures. \ref{fig:H3232}(a) and (b) show the contour plots of 
the field distributions for one unit cell at $H_{32}$.
It is seen that for the s-wave case the contour plot exhibits 
rather square-like shape around the core, which is contrasted 
with a cylindrical shape in the d-wave case. 
Due to the inter-vortex interaction, the magnetic field around 
a vortex tends to extend to the NN vortex direction. 
Then, the field of the $s$-wave case is modified to a square shape, 
reflecting the square vortex lattice shape. 
In the $d_{x^2-y^2}$-wave case, the magnetic field tends to extend 
to $a$ axis and $b$ axis directions due to the gap node 
effect.~\cite{IchiokaDL,Ichioka1,Ichioka2} 
Then, the inter-vortex interaction effect and the gap node effect 
cancel each other, resulting in a cylindrical shape. 

The internal field distributions as a function of the magnetic field, 
which correspond to the resonance frequency distribution in NMR 
experiments, are depicted in Fig. \ref{fig:lineshape}. 
We identify each site (A, B, C, S, and V) in the vortex lattice 
to these distributions.
The one-to-one correspondence between the 
site position and resonance frequency can be used as a guide 
for site-selective NMR experiment: 
Namely, as shown in Figs. \ref{fig:lineshape} (a)
and (b), the NMR signal at the maximum cutoff of the resonance spectrum 
as a function of applied field or probe frequency 
comes from the vortex core at the V-site.
With going away from the center 
(V$\rightarrow$A$\rightarrow$B$\rightarrow$C), the resonance field is 
decreased. 
The signal at the minimum cutoff comes from the C-site. 
The logarithmic singularity of the resonance field comes from the 
saddle point of the field at the S-site.
Thus it is possible to perform the site-selective $T_1({\bf r})$
measurement by tuning the resonance frequency.

\section{Temperature dependence of $T_1$}
\label{sec4}
\subsection{$s$-wave case}

\begin{figure}[tbp]
\begin{center}
\leavevmode
\mbox{
\epsfxsize=60mm
\epsfbox{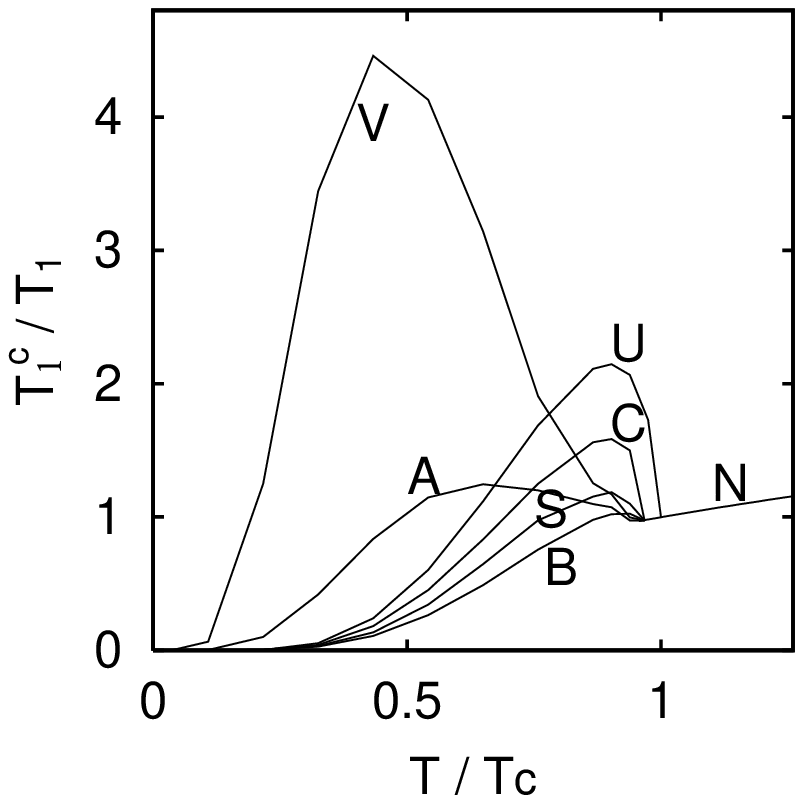}
\epsfxsize=60mm
\epsfbox{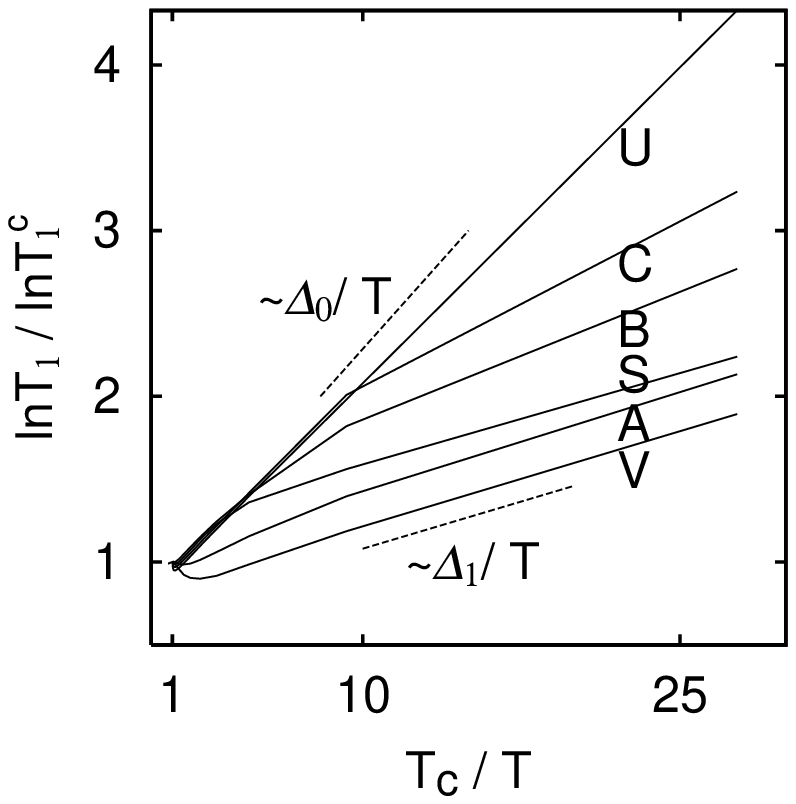}
}
\end{center}
%\vspace{3mm}
\caption{
Temperature dependence of $T_1({\bf r})$ in the $s$-wave case at the 
sites V, A, B, C, and S assigned in Fig. \protect\ref{fig:unit}. 
(a) $T_1(T_c)/T_1(T)$ is plotted as a function of $T/T_c$. 
(b) $\ln T_1(T)/\ln T_1(T_c)$ is plotted as a function of $T_c/T$.
Line U shows the zero field case.
The line N is for the normal state at $T>T_c$. 
}
\label{fig:T1s}
\end{figure}

We now consider the $T$-dependence of $T_1({\bf r})$ at each site defined 
in Fig. \ref{fig:unit}, which reflects the LDOS discussed above.
The $s$-wave case is shown in Fig. \ref{fig:T1s}. 
We plot $T_1({\bf r})^{-1}$ vs. $T$ for each site in Fig. \ref{fig:T1s}(a),  
and re-plot it as $\ln T_1({\bf r})$ vs. $T^{-1}$ 
in Fig. \ref{fig:T1s}(b). 
We also calculate the zero-field case in our formulation. 
It is shown as line U in the figures. 
At the zero field, $T_1 \sim {\rm e}^{\Delta_0 /T}$.
Then, the slope of the $\ln T_1$ vs. $T^{-1}$ plot gives the 
superconducting gap $\Delta_0$, as the line U in Fig.
\ref{fig:T1s}(b).
In the presence of vortices, $T_1$ deviates from the relation ${\rm 
e}^{\Delta_0 /T}$ at low $T$ due to the low-energy excitation around 
the vortex core.
This deviation was reported in the experiments.~\cite{Silbernagel} 
In our results, reflecting the small gap $\Delta_1$ in the $s$-wave 
case [see Figs. \ref{fig:sdos}(a) and (b), also see the V-site 
in Figs. \ref{fig:LDOS}(a)], $T_1$ shows the slope $\Delta_1$ 
at low $T$ in the $\ln T_1$ vs. $T^{-1}$ plot  as seen in 
Fig. \ref{fig:T1s}(b). 
That is, $T_1 \sim {\rm e}^{\Delta_1 /T}$. 
With leaving the vortex center, since the amplitude of the low-energy 
bound states is damped, the weight of ${\rm e}^{\Delta_1 /T}$ 
gradually decreases. 
Then the crossover temperature from ${\rm e}^{\Delta_0 /T}$ to 
${\rm e}^{\Delta_1 /T}$ is lowered.
It is noted that $T_1$ is faster at the S-site than that of the 
B-site, while the S-site is further from the vortex center.
This non-trivial result is due to the vortex lattice effect noted 
above [see the foot denoted by arrows in Fig. \ref{fig:sdos}(a)].
We should also notice the behavior of the coherence peak below 
$T_c$. 
As seen in Fig. \ref{fig:T1s}(a), with approaching the vortex center 
as C$\rightarrow$B, the coherence peak is suppressed. 
But in the vortex core region (lines V and A), a large new peak 
grows at intermediate temperatures. 
This is because the LDOS at the vortex core has peaks at low energy 
$\Delta_1$ instead of the singularity of DOS at $\Delta_0$ as seen 
from Figs. \ref{fig:sdos}(a) and (b).
\subsection{$d$-wave case}

\begin{figure}[tbp]
\begin{center}
\leavevmode
\mbox{
\epsfxsize=60mm
\epsfbox{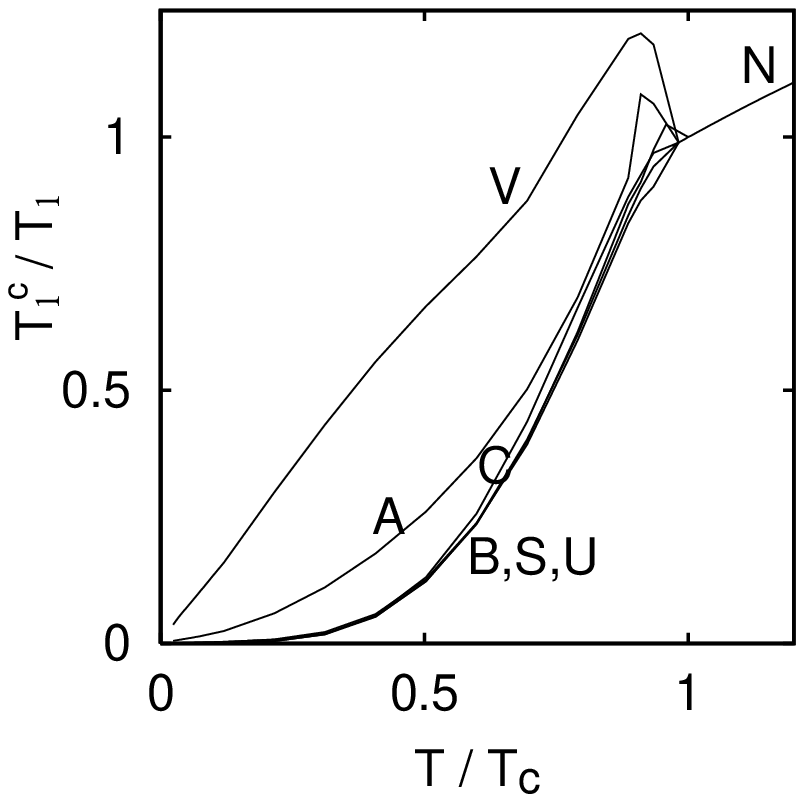}
\epsfxsize=63mm
\epsfbox{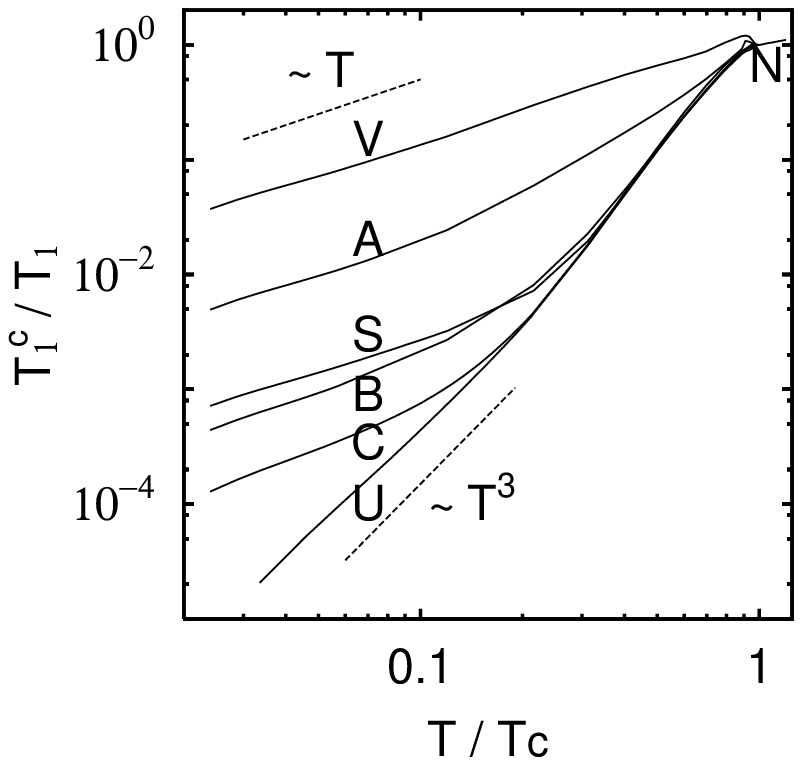}
}
\end{center}
%\vspace{3mm}
\caption{
Temperature dependence of $T_1({\bf r})$ in the $d$-wave case at the 
sites V, A, B, C, and S. 
(a) $T_1(T_c)/T_1(T)$ is plotted as a function of $T/T_c$. 
(b) The log-log plot of (a) is shown. 
Line U shows the zero field case.
The line N is for the normal state at $T>T_c$.
}
\label{fig:T1d}
\end{figure}

As for the $d$-wave case, we plot $T_1({\bf r})^{-1}$ vs. $T$ in Fig.
\ref{fig:T1d}(a), and re-plot it as a log-log plot in Fig.
\ref{fig:T1d}(b).
At zero field (line U), we see the power law relation $T_1^{-1} \sim 
T^3$ of the $d$-wave case as expected.
Note that this can be seen only below $T/T_c \simeq 0.1$ in our case. 
In the presence of vortices, $T_1({\bf r})^{-1}$ deviates from the 
$T^3$-relation, and follows $T_1({\bf r})^{-1} \sim T$ at low 
temperatures.
This deviation was reported in the experiments on high-$T_c$ 
cuprates.~\cite{Ishida} 
The origin of the $T$-linear behavior is the low-energy state 
around vortices in our case, instead of the residual density of 
states due to impurities or defects. 
With approaching the vortex center, the $T$ region of the 
$T$-linear behavior is enlarged and it appears from higher 
temperatures.
As seen in Fig. \ref{fig:LDOS}(b) of the $d$-wave case, the 
superconducting gap is buried by the low-energy state around vortices 
without the small gap of the order $\Delta_0^2/E_{\rm F}$.
Then, $T_1^{-1} \sim T$ at low temperatures in the $d$-wave case 
instead of the relation $T_1 \sim {\rm e}^{\Delta_1 /T}$ in the 
$s$-wave case.
As seen in Fig. \ref{fig:T1d}, $T_1({\bf r})^{-1}$ at the vortex 
center  (line V) is very large compared with the zero-field case 
(line U). 
It reflects the fact that the LDOS of the low-energy state is larger 
than the DOS of the zero-field case as seen in Fig. \ref{fig:LDOS}(b).
Figure \ref{fig:ddos3232} also shows the LDOS at $E\sim 0$ 
normalized by $N(0)$: the DOS in the normal state at the Fermi surface.  
It clearly indicats that the LDOS around the core exceeds
the normal state value.
This short relaxation may be the evidence of the low-energy peak in 
the LDOS by the low-energy core state.

The coherence peak below $T_c$ is taken as a manifestation of 
the $s$-wave symmetry. 
In the $d$-wave case, the coherence peak is absent.  
But in the vortex core region, $T_1^{-1}$ has a peak below $T_c$ even 
in the $d$-wave case. 
We should be careful not to mistake this peak due to the vortex core 
relaxation as the usual coherence peak in the NMR experiment when 
identifying the gap symmetry.

% ###### FIG. 9 start ###################################################
\widetext
%\vspace{3mm}
\narrowtext
% ###### FIG. 9 end ####################################################

% ###### FIG. 10 start ###################################################
\widetext
\vspace{3mm}
\narrowtext
% ###### FIG. 10 end ####################################################

\section{Field dependence of $T_1$}
\label{sec5}

So far, we only consider $T_1(T)$ under a fixed $H(=H_{20})$, 
focusing on the $T$-dependences both
in the $s$-wave and $d$-wave cases.
In this section we investigate the field dependence
of the $T_1(T)$ behavior.

\subsection{Field dependence  of $T_1(T)$}

We calculate the $T$-dependence of $T_1(T)$ for three representative 
fields; the weak field $H_{32}$, the intermediate  field $H_{20}$, 
and the high field $H_{12}$ in order to see how the coherence peak 
evolves and the overall $T$-dependence varies with field.
We show the results for the V-site and the S-site 
(other sites exhibit a similar 
behavior to the S-site, thus not shown here). 
They reflect the field dependence of the LDOS at the vortex 
core and outside of the core.\cite{Wang} 

First, we consider the $s$-wave case. 
It is shown in Fig. \ref{fig:sT1H}. 
As seen in Fig. \ref{fig:sT1H}(a), the $T_1^{-1}$-enhancement at 
the V-site in the intermediate temperature diminishes quickly 
with increasing field. 
This is because the sharp low energy  peaks of $N(E,{\bf r})$  
at the vortex core, which is the origin of the enhancement, is 
smeared by the effect of the quasi-particle transfer  between vortices. 
As shown in  Fig. \ref{fig:sT1H}(b) for the S-site, the coherence 
peak below $T_c$ also diminishes with increasing field. 
In particular, for the high field $H_{12}$, $T_1^{-1}$ shows  
a depression below $T_c$ rather than the enhancement. 
This is caused by the smearing of the superconducting gap by the 
low energy state around the vortex, which extends outside of the 
core region by the quasi-particle transfer between vortices. 
With increasing field, the singularity of the LDOS at the bulk gap edge 
E=$\Delta_0$ is smeared. 
Then the coherence peak diminishes.  
It is qualitatively consistent to the observation on 
${\rm V_3 Sn}$.\cite{Masuda}

\begin{figure}[b]
\begin{center}
\leavevmode
\mbox{
\epsfxsize=60mm
\epsfbox{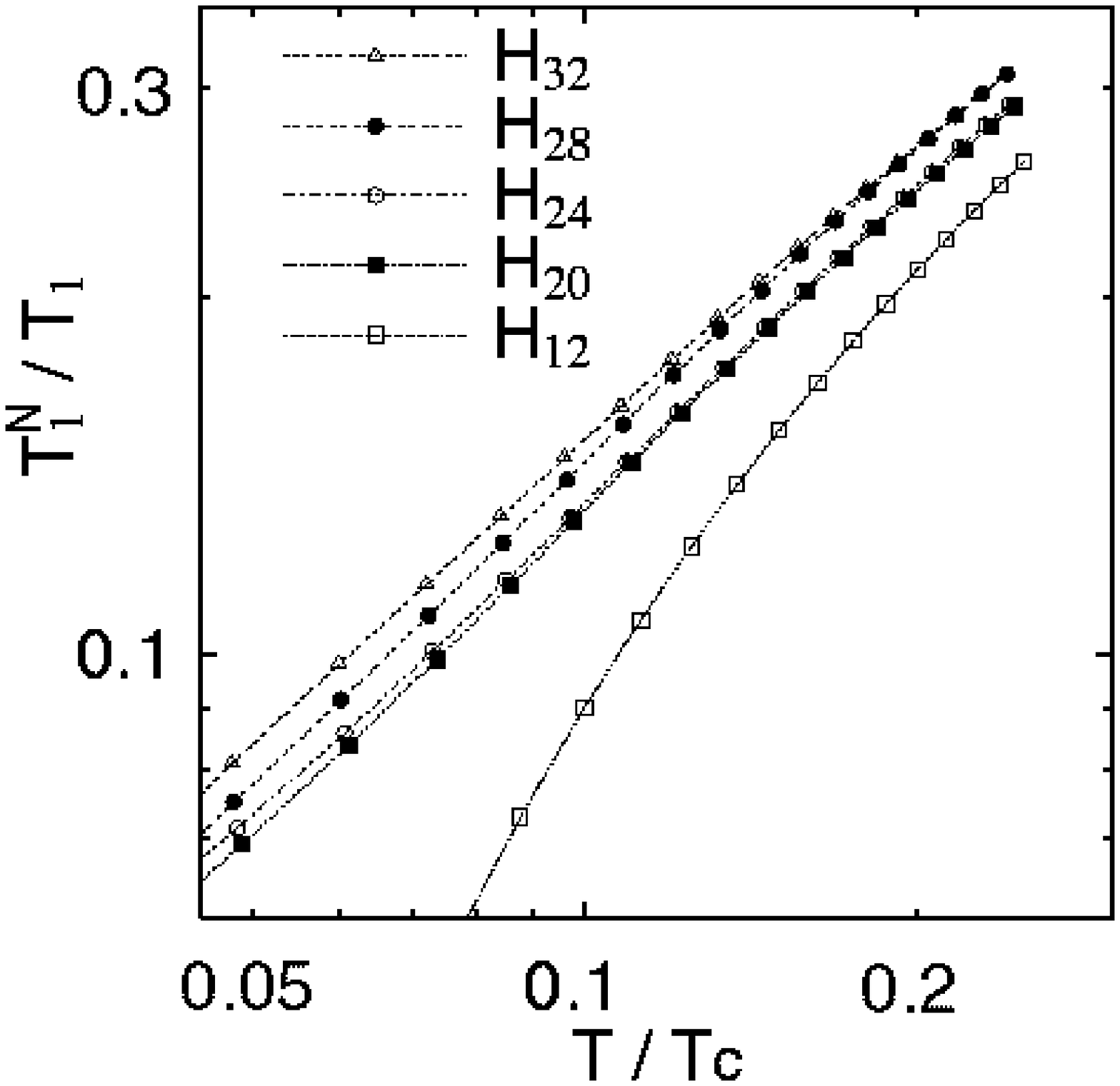}
\epsfxsize=60mm
\epsfbox{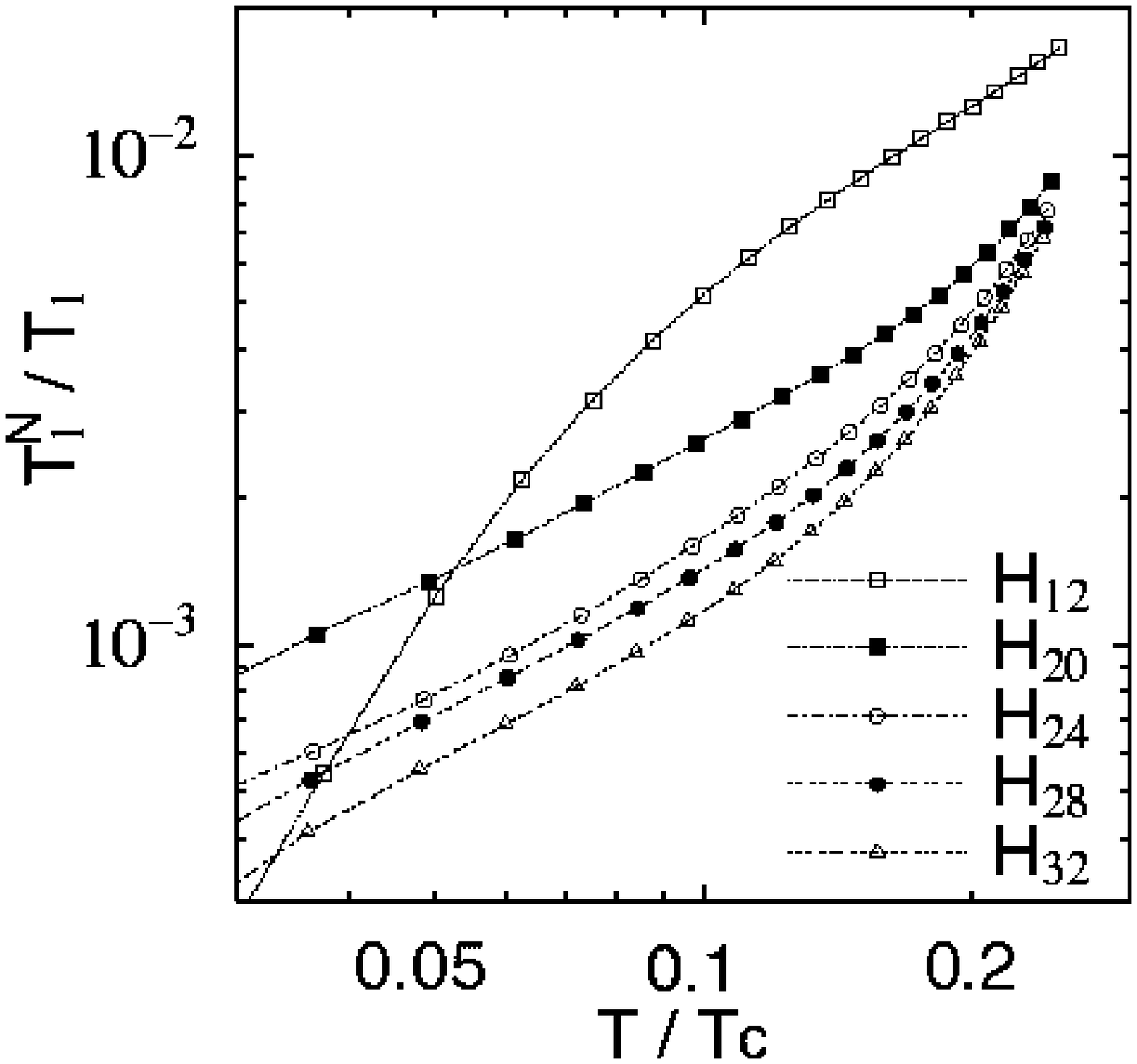}
}
\end{center}
\caption{
The low temperature behavior of $T_1(T)^{-1}$. 
The log-log plots of $T_1(T_c)/T_1(T)$ vs. $T/T_c$ are presented  
for the $d$-wave case at $H_{32}$, $H_{28}$, $H_{24}$, 
$H_{20}$ and the high field $H_{12}$. 
$T_c$ and $T_1(T_c)$ are defined by each $H$. 
(a) the V-site and (b) the S-site. 
}
\label{fig:dT1HL}
\end{figure}

The similar field-dependence of $T_1$ occurs in the $d$-wave case. 
The $T$-dependence of $T_1^{-1}$ is shown in Fig. \ref{fig:dT1HL}. 
At the V-site [Fig. \ref{fig:dT1HL}(a)], the enhancement of 
$T_1^{-1}$ is depressed as $H$ increases. 
It reflects that the low energy peak of the LDOS around the vortex 
is smeared by the quasi-particle transfer between vortices.  
At the S-site [Fig. \ref{fig:dT1HL}(b)],  $T_1^{-1}$ below $T_c$ 
is suppressed as $H$ increases. 
It reflects that the low energy state around the vortex smears the 
$d$-wave superconducting gap-edge. 
The log-log plots of $T_1^{-1}$ vs. $T$ are presented in 
Fig. \ref{fig:dT1HL} to see the low temperature behavior. 
At the V-site [Fig. \ref{fig:dT1HL}(a)], the $T$-linear coefficient 
in $T_1^{-1}(T)$ is depressed with increasing $H$ at low temperature. 
This is because the inter-vortex 
quasi-particle transfer smears the low energy quasi-particle peak 
of the LDOS at the vortex core. 
At the S-site [Fig. \ref{fig:dT1HL}(b)],  
$T_1^{-1}(T)$ is increased with raising $H$ at low temperatures, 
while  $T_1^{-1}(T)$ is decreased with raising $H$ near $T_c$ 
[Fig. \ref{fig:dT1HL}(b)].  
Because the vortex contribution is increased and the amplitude of the 
low energy state extending outside the vortex core becomes large, 
the relaxation time at low temperatures becomes short 
with increasing external magnetic field.  
This tendency coincides qualitatively
with the observation of high-$T_c$ cuprates by Ishida 
{\it et al.}~\cite{Ishida} or an organic superconductor 
$\kappa$-(ET)$_2$Cu[N(CN)$_2$]Br by Mayaffre {\it et al.}~\cite{Mayaffre} 

We note that the quantum oscillations due to the Landau band 
quantization affects the low temperature behavior of $T_1$ at 
extreme high fields. 
In Fig. \ref{fig:dT1HL}, $T_1^{-1}(T)$ at the high field $H_{12}$ 
deviates from the $T$-linear behavior. 
It deviates downward (upward), when the LDOS is minimum (maximum) 
at the Fermi energy in the quantum oscillation of the LDOS 
(Fig. \ref{fig:ddoshigh}). 
% ###### FIG. 11 start ###################################################
\widetext
\begin{figure}[tbp]
\begin{center}
\leavevmode
\mbox{
\epsfxsize=60mm
\epsfbox{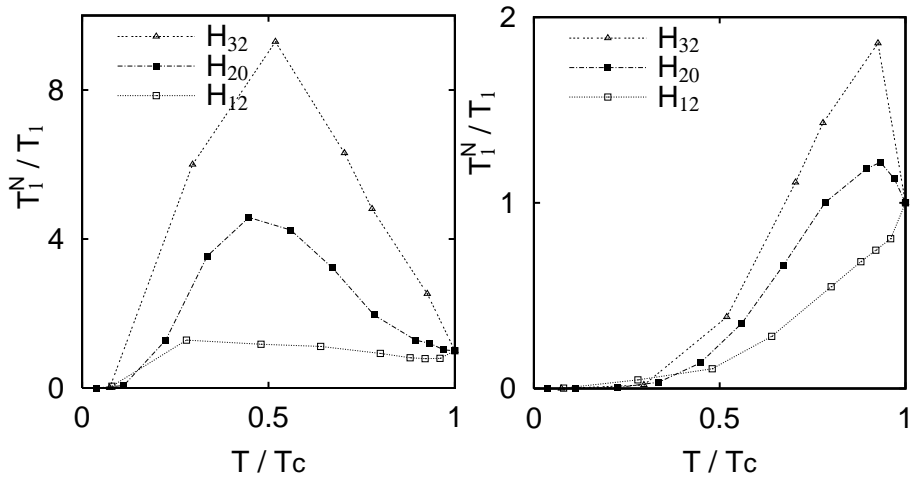}
\epsfxsize=60mm
\epsfbox{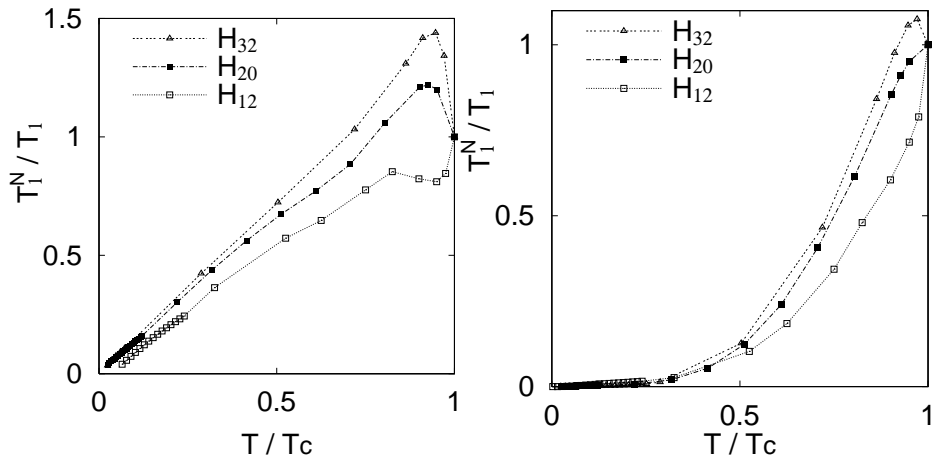}
}
\end{center}
\caption{
$T_1(T_c)/T_1(T)$ is plotted as a function of $T/T_c$ for 
the $s$-wave case at $H_{32}$, $H_{20}$, and $H_{12}$. 
$T_c$ and $T_1(T_c)$ are defined by each $H$. 
(a) the V-site and (b) the S-site.
}
\label{fig:sT1H}
\end{figure}
\narrowtext
% ###### FIG. 11 end ####################################################
% ###### FIG. 12 start ###################################################
\widetext
\vspace{5.5mm}
\begin{figure}[tbp]
\begin{center}
\leavevmode
\mbox{
\epsfxsize=60mm
\epsfbox{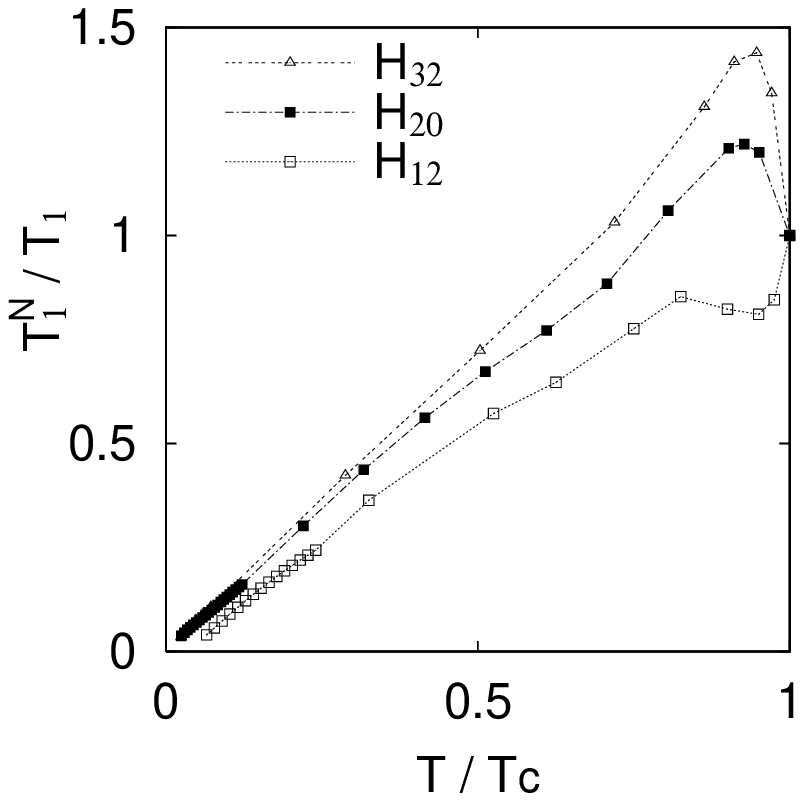}
\epsfxsize=63mm
\epsfbox{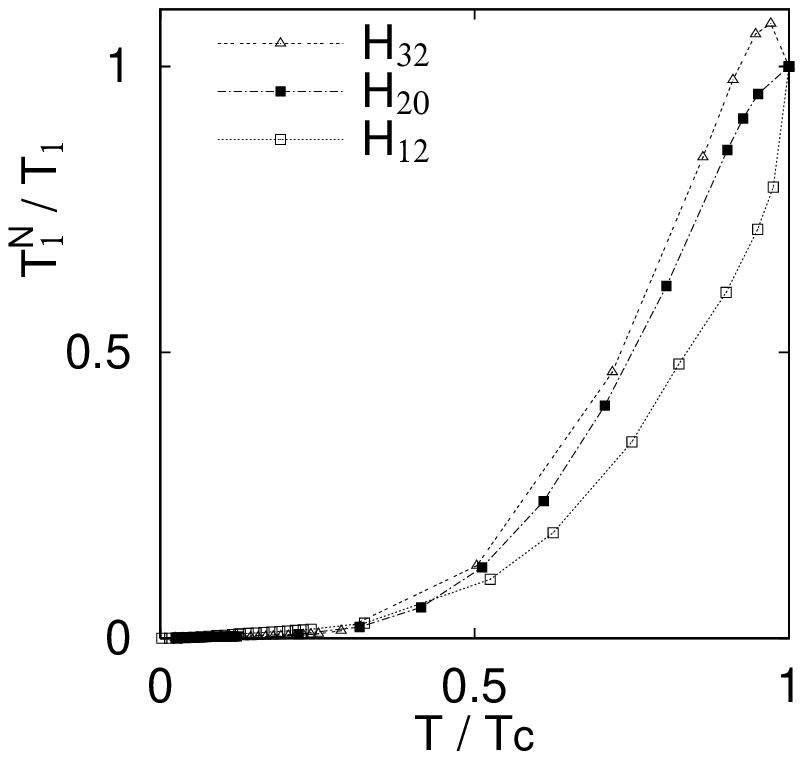}
}
\end{center}\caption{
$T_1(T_c)/T_1(T)$ is plotted as a function of $T/T_c$ for 
the $s$-wave case at $H_{32}$, $H_{20}$, and $H_{12}$. 
$T_c$ and $T_1(T_c)$ are defined by each $H$. 
(a) the V-site and (b) the S-site.
}
\label{fig:sT1H}
\end{figure}
\narrowtext
% ###### FIG. 12 end ####################################################
% ###### FIG. 13 start ###################################################
\widetext
\vspace{5.5mm}
\narrowtext
% ###### FIG. 13 end ####################################################

\subsection{$T_1({\bf r},T)$ vs. LDOS}

\begin{figure}[tbp]
\epsfxsize=75mm
\epsfbox{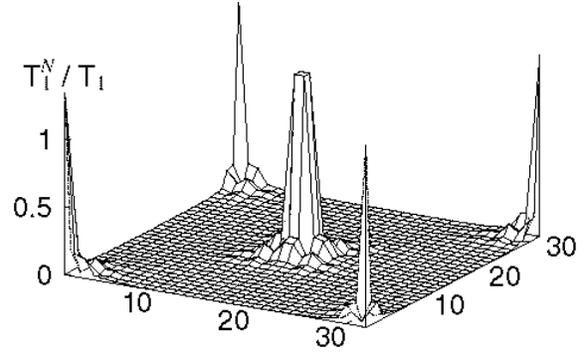}
\caption{
Topographic view of the site-dependent relaxation time 
$T_1^N/T_1({\bf r})$ for the $d$-wave case at $H_{32}$ and $T \sim 0$. 
It is normalized by its normal state value $T_1^N$. 
One unit cell (32$\times$32 atomic sites) is shown.
The vortices are located at the center and corners of the figure.
}
\label{fig:T13232}
\end{figure}

Here we investigate the relationship between 
$T_1^{-1}({\bf r})$ at low temperatures and the low energy
excitations $N(E=0,{\bf r})$ in the $d$-wave case.
In the low temperature region, $T_1^{-1}$ shows $T$-linear behavior. 
First, we plot the spatial distribution of 
$T_1^{-1}({\bf r},T\sim 0)$ at $H_{32}$ for a unit cell in 
Fig. \ref{fig:T13232}. 
This shows that $T_1^{-1}({\bf r},T\sim 0)$ normalized 
by the normal state value $T_1^N$ is largest at the 
core and exceeds its normal state value.
Namely, with approaching the vortex core,
$T_1^{-1}$ increases. This tendency is in qualitative agreement 
with a measurement by Milling and Slichter\cite{milling} 
on $^{63}$Cu NMR experiment of YBCO.

\begin{figure}[tbp]
\begin{center}
\leavevmode
\mbox{
\epsfxsize=60mm
\epsfbox{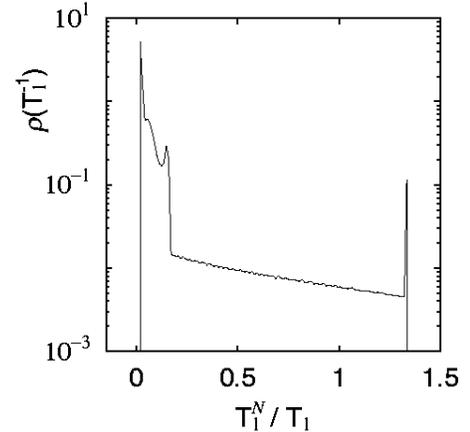}
}
\end{center}
\caption{
Histogram $\rho(T_1^{-1})$ of $T_1^{-1}$ corresponding to
Fig. \ref{fig:T13232}. 
}
\label{fig:histo}
\end{figure}

The histogram $\rho(T_1^{-1})$ of $T_1^{-1}({\bf r},T\sim 0)$ 
is shown in Fig. \ref{fig:histo}, 
where the values of $T_1^{-1}({\bf r},T\sim 0)$ for the
total 32$\times$32 sites is classified according to its magnitudes. 
This indicates that the vast majority sites outside the vortex core 
exhibit long relaxation times, but certain few sites around the vortex 
core exhibit short relaxation time, which contains useful information 
on the site-dependent low energy  excitations associated with the vortex. 

\begin{figure}[tbp]
\begin{center}
\leavevmode
\mbox{
\epsfxsize=60mm
\epsfbox{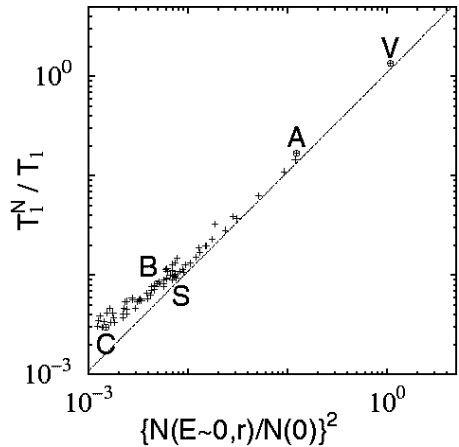}
}
\end{center}
\caption{
The linear relationship between $T_1^N/T_1({\bf r})$
and $[N(E=0,{\bf r})/N(0)]^2$. 
We plot the $32 \times 32$ lattice point values of the normalized 
$T_1^N/T_1({\bf r})$ shown in Fig. \ref{fig:T13232} and 
the square of the normalized LDOS $N(E=0,{\bf r})/N(0)$ 
shown in Fig. \ref{fig:ddos3232}. 
The letters in the figures denote the sites defined 
in Fig. \ref{fig:unit}. 
The dotted line shows the linear relation 
$T_1^N/T_1({\bf r}) \propto [N(E=0,{\bf r})/N(0)]^2$.
}
\label{fig:T1N0}
\end{figure}

This topography of $T_1^{-1}({\bf r},T\sim 0)$
in Fig. \ref{fig:T13232}
looks similar to that of LDOS $N(E=0,{\bf r})$
in Fig. \ref{fig:ddos3232}.
In fact, as shown in Fig. \ref{fig:T1N0} 
where $T_1^{-1}({\bf r},T\sim 0)$ vs. $N^2(E=0,{\bf r})/N^2(0)$ 
is plotted, the linear relationship between them is apparently seen. 
It implies that $T_1^{-1}({\bf r},T\sim 0)$ at low $T$ is 
governed by the low energy excitations at each site. 
Then, the site-selective $T_1^{-1}({\bf r},T\sim 0)$ is a 
good measure of the local core excitations.
The LDOS around the vortex core can be estimated quantitatively 
from the slope of $T_1^{-1}({\bf r})$ at low $T$.

% ###### FIG. 14 start ###################################################
%\vspace{5.5mm}
% ###### FIG. 14 end ####################################################
% ###### FIG. 15 start ###################################################
\vspace{5.5mm}
% ###### FIG. 15 end ####################################################
\widetext
\narrowtext

% ###### FIG. 16 start ###################################################
% ###### FIG. 16 end ####################################################

\section{Conclusion and discussions}
\label{sec6}

We have calculated the nuclear relaxation time 
$T_1$ in mixed state in the microscopic framework of the BdG theory 
and demonstrated that the site-selective $T_1$ is a good probe to 
extract detailed information on low lying vortex core excitations. 
We have investigated the site-dependence, $T$-dependence and 
field-dependence of $T_1$ both for the $s$-wave 
and $d$-wave superconductors.

Traditionally, the vortex contribution was considered as the spin 
diffusion to the normal region of the vortex 
core,~\cite{Ishida,Caroli} and $T_1$ is treated as the spatial average.
However, we can investigate the position dependence of $T_1({\bf r})$ 
around vortices through the resonance field dependence.
This is an advantage of NMR over other methods. 
We should clarify the local mechanism of the relaxation (i.e., 
whether the relaxation occurs locally, or it is averaged by the spin 
diffusion). 
It is noted that in the clean limit the vortex core region is 
not a simple core filled by normal state 
electrons.~\cite{machida,Hayashi,Hayashi2,Volovik,IchiokaDL,IchiokaDS,IchiokaSL,Ichioka1,Ichioka2,Ichioka4,Ichioka5} 
There, the characteristic $T$-dependence is expected near the vortex 
core  other than a simple $T$-linear behavior, reflecting the rich 
structure of the low energy state around the vortex core. 

We expect that the NMR imaging study  just explained here 
will provide vital information for the vortex core state 
in high-$T_c$ cuprates.
As for the problem whether the quantization of the energy levels 
occurs or not, $T_1 \sim {\rm e}^{\Delta_1/T}$ if the gap $\Delta_1$ 
($\sim \Delta_0^2/E_{\rm F}$) is present in the excitation due to the 
quantization.
If this small gap is absent, $T_1^{-1} \sim T$. 
As for the problem whether the zero-energy peak exists or not in the 
core state, the relaxation at the core becomes eminently faster than 
that of the zero-field case (or that far from the vortex) at low 
temperature, if the zero-energy peak exists in the LDOS as suggested 
in the theoretical study.
If the peak structure is absent within $\Delta_0$ as reported in the 
STM experiments on BSCCO, the relaxation is slow even at the vortex 
core as in the zero-field superconducting case.

We have pointed out a pitfall of the conventional
procedure of the NMR experiment to extract the nodal gap structure 
by analyzing the power law of the $T$-dependence 
of $T_1$ or the coherence peak. 
Even in the $d$-wave pairing,  $T_1^{-1}$ shows enhancement 
below $T_c$ by the vortex core contribution. 

Finally, we stress that, to perform an idealized site-selective 
NMR measurement, we need a clear NMR resonance line shape of 
the vortex lattice as shown in Fig. \ref{fig:lineshape} or 
as that obtained by a beautiful experiment on 
vanadium.\cite{Fite,Kung}  
In the resonance line shape, the signal near the upper (lower) 
cutoff of the field distribution gives the information from 
the vortex core (the outside of the core). 
However, even in the case when the resonance line shape of 
the vortex lattice is not clear, if we analyze the fast and slow 
relaxation processes separately, the fast relaxation process 
includes the information of the vortex core 
contribution.~\cite{Mukuda} 
We hope that future NMR experiments confirm our proposal of 
the site-selective NMR, and clarify the relation of $T_1({\bf r})$ 
and $N(E,{\bf r})$. 
If this experimental method is established, 
it can be a powerful method to investigate the exotic mechanism of 
the unconventional superconductors by spatially imaging the 
low energy quasi-particle excitation around the vortex cores.

\newpage
%%%%%%%%%%%%%%%%%%%%%%%%%%%%%%%
\widetext
\appendix
\section{}
\label{secA}

The equation of motion for the Green function in a 2$\times$2 matrix 
form is written in the coordinate space as 
%%%
\begin{equation}
\sum_{{\bf x''}}
\{
-\hbar{\partial\over{\partial \tau}}{\bf 1}\delta({\bf x}-{\bf x''})-
\left(
\begin{array}{cc}
K({\bf x},{\bf x''})&\Delta({\bf x},{\bf x''})\\ 
\Delta^\dagger({\bf x},{\bf x''})&-K^\ast({\bf x},{\bf x''})
\end{array} 
\right)
\}
{\hat g}({\bf x''} \tau,{\bf x'} \tau')
=\hbar\delta({\bf x}-{\bf x'})
\delta({\tau}-{\tau'}){\bf 1}
\label{eq:eqgreen}
\end{equation}
%%%
where ${\bf 1}$ is the 2$\times$2 unit matrix.
We put the completeness relation:
%%%
\begin{equation}
\delta({\bf x}-{\bf x'}){\bf 1}
=\sum_{E_{\alpha}>0}
\left(
\begin{array}{cc} 
u_{\alpha}({\bf x})&-v^\ast_{\alpha}({\bf x})\\
v_{\alpha}({\bf x})&u^\ast_{\alpha}({\bf x})
\end{array}
\right)
\left(
\begin{array}{cc} 
u^\ast_{\alpha}({\bf x'})&v^\ast_{\alpha}({\bf x'})\\
-v_{\alpha}({\bf x'})&u_{\alpha}({\bf x'})
\end{array}
\right) 
\label{eq:eqgreen2}
\end{equation}
%%%
before ${\hat g}({\bf x''} \tau,{\bf x'} \tau')$ in Eq. (\ref{eq:eqgreen}), 
and use the BdG equation of the form 
%%%
\begin{equation}
\sum_{\bf x'}
\left(
\begin{array}{cc}
K({\bf x},{\bf x'})&\Delta({\bf x},{\bf x'})\\ 
\Delta^\dagger({\bf x},{\bf x'})&-K^\ast({\bf x},{\bf x'})
\end{array} 
\right)
\left(
\begin{array}{cc} 
u_{\alpha}({\bf x'})&-v^\ast_{\alpha}({\bf x'})\\
v_{\alpha}({\bf x'})& u^\ast_{\alpha}({\bf x'})
\end{array}
\right)
=
\left(
\begin{array}{cc} 
u_{\alpha}({\bf x})&-v^\ast_{\alpha}({\bf x})\\
v_{\alpha}({\bf x})&u^\ast_{\alpha}({\bf x})
\end{array}
\right)
\left(
\begin{array}{cc} 
E_{\alpha} & 0\\ 0 & -E_{\alpha} 
\end{array}
\right)
\label{eq:BdG22}
\end{equation}
%%%
for $E_\alpha >0$. 
After the Fourier transformation of $\tau$ to $\omega_n$, 
the inversion of the matrix product leads to 
%%%
\begin{equation}
{\hat g}({\bf x},{\bf x'},\omega_n)
=\sum_{E_{\alpha}>0}
\left(
\begin{array}{cc} 
u_{\alpha}({\bf x})&-v^\ast_{\alpha}({\bf x})\\
v_{\alpha}({\bf x})&u^\ast_{\alpha}({\bf x})
\end{array}
\right)
\left(
\begin{array}{cc} 
\frac{1}{i\omega_n-E_{\alpha}/\hbar} & 0\\
0 & \frac{1}{i\omega_n+E_{\alpha}/\hbar} 
\end{array}
\right)
\left(
\begin{array}{cc} 
u^\ast_{\alpha}({\bf x'})&v^\ast_{\alpha}({\bf x'})\\
-v_{\alpha}({\bf x'})&u_{\alpha}({\bf x'})
\end{array}
\right)
\label{eq:eqgreen5}
\end{equation}
%%%
with the help of the completeness and the orthogonal relation of 
the wave functions. 
In the term with $(i\omega_n+E_{\alpha}/\hbar)^{-1}$, 
we use the symmetry property: 
$u_\alpha \rightarrow -v^\ast_\alpha$, 
$v_\alpha \rightarrow u^\ast_\alpha$ 
when $E_\alpha \rightarrow -E_\alpha$. 
Then, Eqs. (\ref{eq:green2-1})-(\ref{eq:green2-4}) are obtained.

%%%%%%%%%%%%%%%%%%%%%%%%%%%%
\section{}
\label{secB} 

The spin-spin correlation function is defined as
%%%
\begin{equation}
\chi_{-,+}(x,x')=\left\langle T_{\tau}[S_-(x)S_+(x')] \right\rangle 
=\left\langle T_{\tau} [
\psi^\dagger_{\downarrow}(x) \psi_{\uparrow}(x) 
\psi^\dagger_{\uparrow}(x')  \psi_{\downarrow}(x')
] \right\rangle 
\end{equation}
%%%
with $x\equiv({\bf x},\tau)$. 
This can be rewritten in terms of the thermal Green functions, 
%%%
\begin{equation}
\chi_{-,+}(x,x')=
 g_{11}(x,x')g_{22}(x,x')-g_{12}(x,x')g_{21}(x,x') . 
\label{eq:chi2}
\end{equation}
%%%%%
After the Fourier transformation of $\tau$, we obtain
%%%
\begin{equation}
\chi_{-,+}({\bf x},{\bf x'},\Omega_n)=T\sum_{\omega_n}
\{ g_{11}({\bf x},{\bf x'},\omega_n)
   g_{22}({\bf x},{\bf x'},\Omega_n-\omega_n)
  -g_{12}({\bf x},{\bf x'},\omega_n) 
   g_{21}({\bf x},{\bf x'},\Omega_n-\omega_n) \}.
\label{eq:chi3}
\end{equation}
%%%%%%%%%%%%%%
Then, substituting Eqs. (\ref{eq:green2-1})-(\ref{eq:green2-4}) 
and performing $\omega_n$-sum, the spin-spin correlation function 
with real frequency $\Omega$ is reduced to
%%%
\begin{equation}
\chi_{-,+}({\bf x},{\bf x'},i\Omega_n\rightarrow\Omega+i\eta)=
-\sum_{\alpha,\beta}
u_{\alpha}({\bf x})u^{\ast}_{\beta}({\bf x})
\{
u^{\ast}_{\alpha}({\bf x'})u_{\beta}({\bf x'})
+v^{\ast}_{\alpha}({\bf x'})v_{\beta}({\bf x'})
\}
{{f(E_{\alpha})-f(E_{\beta})}\over{E_{\alpha}-E_{\beta}-\Omega-i\eta}}.
\label{eq:chi}
\end{equation}
%%%%%%%%%%%%%
Here, we used the symmetry property: 
$u_\alpha \rightarrow -v^\ast_\alpha$, 
$v_\alpha \rightarrow u^\ast_\alpha$ 
when $E_\alpha \rightarrow -E_\alpha$. 

\newpage
\narrowtext

%%%% references %%%%%%%%%%%%%%%%%%%%%%%%%%%%%%%%%%%%%%%

%%%%%%%%%%%%%%%%%%%%
\newpage
%%%%%%%%%%%%%%%%%%%%
\widetext
%%%%%%%%%%%%%%%%%%%%%%%%%%%%%%%%%%%%%%%%%%%%%%%%%%%%%%%
\end{document}